\documentstyle[11pt,amssymb]{article}

\makeatletter
\@addtoreset{equation}{section}
\makeatother

\def\dalemb#1#2{{\vbox{\hrule height .#2pt
        \hbox{\vrule width.#2pt height#1pt \kern#1pt
                \vrule width.#2pt}
        \hrule height.#2pt}}}

\def\cN{{\cal N}}
\def\cA{{\cal A}}

\def\cL{{\cal L}}

\def\ca{{\cal I}}
\def\caj{{\cal J}}

\def\0{{\sst{(0)}}}
\def\1{{\sst{(1)}}}
\def\2{{\sst{(2)}}}
\def\3{{\sst{(3)}}}
\def\4{{\sst{(4)}}}
\def\5{{\sst{(5)}}}
\def\6{{\sst{(6)}}}
\def\7{{\sst{(7)}}}
\def\8{{\sst{(8)}}}

\def\ep{\epsilon}
\def\td{\tilde}

\def\half{{\textstyle{1\over2}}}
\def\hp{ \frac{1}{2}}
\def\qu{{\textstyle{1\over 4}}}

\let\a=\alpha  \let\g=\gamma \let\d=\delta \let\e=\epsilon
  \let\q=\theta  \let\k=\kappa
\let\l=\lambda \let\m=\mu   
\let\s=\sigma \let\t=\tau  \let\f=\phi  
\let\w=\omega  \let\D=\Delta  \let\L=\Lambda
    
    \let\G=\Gamma
\let\la=\label  
  
\def\nn{\nonumber} \def\bd{\begin{document}} \def\ed{\end{document}}
\def\ds{\documentstyle} \let\fr=\frac \let\bl=\bigl \let\br=\bigr
\let\Br=\Bigr \let\Bl=\Bigl
\let\bm=\bibitem
\let\na=\nabla
\let\pa=\partial \let\ov=\overline
\newcommand{\be}{\begin{equation}}
\newcommand{\ee}{\end{equation}}
\def\ba{\begin{array}}
\def\ea{\end{array}}
\def\ft#1#2{{\textstyle{{\scriptstyle #1}\over {\scriptstyle #2}}}}
\def\fft#1#2{{#1 \over #2}}
\def\del{\partial}
\def\sst#1{{\scriptscriptstyle #1}}
 \def\oneone{\rlap 1\mkern4mu{\rm l}}
\def\ie{{\it i.e.\ }}
\def\via{{\it via}}
\def\semi{{\ltimes}}
\def\str{{\rm str}}
\def\Dm{{{D_{\sst{max}}}}}
\def\vac{ \left | 0 \right \rangle }
\def\kvac{ \left | k \right \rangle }

\def\sp{\; \; \;}

\def\bol{ \left | B (p^+) \right \rangle}
\def\bo1{ \left | B^0 (p^+) \right \rangle}

\def\bolt{ \left | B (p^+) \right \rangle_{\t}}

\def\boxl{ \left | B (x^-) \right \rangle}

\def\barray{ \left | B_{\rm{array}} \right \rangle}

\newcommand{\hsp}{\hspace{0.5cm}}

\newcommand{\ho}[1]{$\, ^{#1}$}
\newcommand{\hoch}[1]{$\, ^{#1}$}
\newcommand{\bea}{\begin{eqnarray}}
\newcommand{\eea}{\end{eqnarray}}
\newcommand{\ra}{\rightarrow}
\newcommand{\lra}{\longrightarrow}
\newcommand{\Lra}{\Leftrightarrow}
\newcommand{\ap}{\alpha^\prime}
\newcommand{\bp}{\tilde \beta^\prime}
\newcommand{\tr}{{\rm tr} }
\newcommand{\Tr}{{\rm Tr} }
\newcommand{\NP}{Nucl. Phys. }

\newcommand{\spin}{{\it Spinoza Institute, University of Utrecht,\\
Postbus 80.195, 3508 TD Utrecht, The Netherlands}\\
{\tt email:taylor@phys.uu.nl}}
\newcommand{\ams}{{\it Institute for Theoretical Physics,
University of Amsterdam, \\
Valckenierstraat 65, 1018XE Amsterdam, The Netherlands} \\
{\tt email:skenderi@science.uva.nl}}
\newcommand{\auth}{{\large Kostas Skenderis\hoch{\star} and
Marika Taylor\hoch{\dagger}}}

\begin{document}
\begin{flushright}
\hfill{\bf hep-th/0311079}\\
\hfill{SPIN-2003/29} \\
\hfill{ITF-2003/48} \\
\hfill{ITFA-2003-47}
\end{flushright}

\vspace{15pt}

\begin{center}

{\Large \bf Properties of branes in curved spacetimes}

\vspace{20pt}

\auth
\vspace{15pt}

\vspace{8pt}

{\hoch\star \ams}

\vspace{8pt}

{\hoch\dagger \spin}

\vspace{15pt}

\underline{ABSTRACT}
\end{center}
A generic property of curved manifolds is the existence of focal points. 
We show that branes located at focal points of the geometry satisfy
special properties. Examples of backgrounds to which our discussion
applies are $AdS_m \times S^n$ and plane wave backgrounds. 
As an example, we show that a pair of $AdS_2$ branes located at 
the north and south pole of the $S^5$ in $AdS_5 \times S^5$ are 
half supersymmetric and that they are dual to a two-monopole
solution of ${\cal N}=4$ SU(N) SYM theory. Our second example
involves spacelike branes in the (Lorentzian) plane wave. 
We develop a modified lightcone gauge for the open
string channel, analyze in detail the cylinder diagram
and establish open-closed duality. In the new gauge the open 
string feels an inverted harmonic oscillator potential. 
When the branes are located at focal points of the geometry the amplitude 
acquires most of the characteristics of flat space amplitudes.
In the open string channel the special properties are due
to stringy modes that become massless.

\noindent

\pagebreak
\setcounter{page}{1}

\tableofcontents

\section{Introduction}

Understanding branes on curved backgrounds is an important
problem. One of the most basic properties that one would like
to understand is the interactions between branes and of branes
with closed strings. In string perturbation 
theory the leading interaction is given by the cylinder diagram. 
Studying this diagram and its properties is important for a
variety of reasons. Firstly, it can be viewed as either
a one-loop open string diagram or as tree level
exchange of closed strings. Showing that the two
descriptions yield the same answer provides a non-trivial
consistency check, and furthermore, this equivalence
is the prototype example of a gauge theory/gravity
duality. Secondly, possible divergences in the amplitude
signal physical effects. For instance, in some cases cancelling
such one-loop divergences leads to string-loop corrections
to the beta function equations (and consequently corrections to the
background) \cite{Callan:1987px,Polchinski:1987tu,Callan:1988wz}.
In other cases the divergence is a signal of an instability:
a finite amplitude is obtained via analytic continuation but it has
an imaginary part. The latter yields the decay rate of the brane.

In flat spacetime the cylinder amplitude between two parallel branes
exhibits the following behavior: 
(i) it is proportional to the volume of the D-branes; 
(ii) it vanishes for any separation if the branes are supersymmetric;
(iii) the supersymmetries that annihilate the corresponding
boundary state commute with time evolution. 
The first property is due to translational invariance of the
interactions. Because of this factor the amplitude
for infinitely extended branes diverges,
but this divergence does not signify an instability or the onset of
string-loop corrections. The second property is due to fermionic zero
modes. The last property is dictated by the flat space superalgebra.

In curved spacetime none of these properties are expected to
hold in general. For instance, even if individual branes
are translationally invariant along their worldvolume, the
system of a pair of branes will not in general retain this
property. As a result the cylinder amplitude will not automatically
be proportional to the worldvolume of the branes, and it becomes
a non-trivial task to disentangle divergences that are due to the infinite
volume of the brane from the physical divergences. One purpose of this
work is analyze general properties of branes in curved spacetimes
that would allow the better understanding of the meaning of the amplitudes.

In a general curved manifold we expect that special features appear
when the branes are located at focal points of the background geometry.
In these cases the system acquires new continuous zero modes: these
are associated with open strings with ends on the two D-branes
that lie along geodesics. Since any geodesic leaving from the original
brane ends up on the other branes, there are new zero modes, namely the
modes that parametrize the different geodesics. For the Neumann directions
these are the position and velocity of the string ends, and for Dirichlet
directions the position and velocity of $\pa_\s X^{r'}$ at the end of
the string. In particular, the amplitudes will again be proportional
to the volume of the branes since we have regained translational
invariance along the worldvolume directions. Furthermore, if the
branes are target space supersymmetric, supersymmetry
will imply new fermionic zero modes, namely the fermionic partners
for the new bosonic zero modes.
Because of these modes, the cylinder amplitude is expected to vanish
and we thus find that these amplitudes behave as in flat spacetime.

Focal points are a generic feature of curved manifolds.
Examples include many of the backgrounds that enter in 
gravity/gauge theory dualities. Perhaps the prototype examples of 
curved manifolds with focal points are spheres: all geodesics that 
leave the north pole reconverge at the south pole.
AdS spacetime itself has spacelike focusing points: timelike geodesics 
reconverge after global time $\pi$.  Thus our discussion is relevant 
for branes in all backgrounds that involve $AdS$ and/or spheres.
In many of these cases we do not yet know how to solve string
theory, so computing the cylinder diagram is out of reach.
The previous discussion however implies that there are supersymmetric
configurations that involve branes located at focal points of 
the geometry. This in turn implies via the gravity/gauge 
theory duality that there must exist dual 
supersymmetric configurations on the gauge theory side. 
We thus obtain an additional set of configurations that 
one should match between the two sides of the duality.
 
To illustrate this discussion we analyze $AdS_2$ branes on $AdS_5 \times S^5$.
We have previously shown \cite{ST} that $AdS_2$ branes wrapping the 
time and radial coordinate of $AdS_5$ preserve 16 supercharges.
Here we observe that a system of two such branes, one at the north pole and 
another at the south pole preserve the same 
number of supercharges. We then show that this configuration 
corresponds to a two monopole solution of the ${\cal N}=4$ SU(N) SYM theory,
which is also a half supersymmetric configuration.

Another set of examples of curved manifolds with focal points are 
plane waves. A particularly interesting case is the maximally 
supersymmetric background of IIB supergravity as it is 
the Penrose limit of $AdS_5 \times S^5$. The focal points
of the geometry descend from corresponding focal points 
on $AdS_5 \times S^5$ : geodesics that reconverge 
on {\it both} $AdS_5$ and the circle of $S^5$ along which we boost
are part of the resulting plane wave spacetime. 
We thus expect that a pair of spacelike branes in the (Lorentzian)
plane wave exhibits special properties when located 
at the focal points. 

This example has the advantage that string theory on this
background is solvable in lightcone gauge and thus the cylinder 
diagram can be computed. In fact the computation
of the cylinder diagram in the closed string channel
for the branes of interest here was carried out in \cite{BGG} and 
(as expected) none of the above mentioned flat space properties hold. 
Moreover, as we discuss in detail, the corresponding amplitudes are 
generically not just non-zero, they are infinite. Understanding 
the meaning of these infinities was one of the motivations of this
work. The amplitude, however, recovers the flat space characteristics 
when the branes are located at focal points. In these cases, the 
infinities of the amplitudes are just due to the infinite volume 
of the branes. We expect that the infinities at generic
separations are also of the same nature, even though they 
cannot be directly expressed as volume divergences.

Open strings in the standard lightcone gauge can only describe 
timelike branes that wrap both lightcone directions. In order 
to be able to analyze the open-closed duality we are led to
develop a modified lightcone gauge where $X^+ \sim \s$.  Recall that 
the worldsheet theory for closed strings in the standard lightcone 
gauge describes bosons and  fermions in a harmonic oscillator potential.
The open string theory however in the modified lightcone
gauge describes open strings in an {\it inverted} harmonic oscillator
potential. Throughout our analysis the worldsheet theory is 
Lorentzian.

The special separations that on the closed string side correspond to 
focal points are mapped under open-closed duality to a specific
value of the mass for the open string. As we approach this value of the mass 
one of the stringy modes becomes massless, and the special properties
that were due to the infinite number of geodesics in the closed string 
channel are now due to the presence of extra massless modes.

On a generic background, the cylinder diagram between
two supersymmetric branes which are related by symmetry transformations is
not necessarily zero. This can even be the case for branes which are
parallel, i.e. separated using a translational symmetry, if
such translations do not commute with the supersymmetries. 

Again these properties are nicely illustrated by the example of
spacelike branes in the plane wave. 
In this case the relevant translational symmetry is the lightcone Hamiltonian
$H$ which does not commute with target space
supersymmetry. So if the boundary state $|B\rangle_0$ (defined at $x^+=0$ )
is annihilated by a combination of supercharges, the time-evolved state
$|B,x^+ \rangle {=} e^{-i H x^+} |B\rangle_0$,
will generically be annihilated by a different set of supercharges. 
As a result
${}_0\langle B| e^{-i H x^+} |B \rangle_0$ will in general be non-zero.
However, exactly when the time of evolution is such that 
the geodesic focal point 
is reached, the set of supercharges that annihilates 
the boundary state rotates back to the original set. As a result
the cylinder amplitude for two branes at this separation vanishes.

The physical relevance of the spacelike branes discussed here is
unclear, since they exhibit known problematic features of 
spacelike branes, for instance, they source imaginary fluxes. 
Actually, as we discuss,
these branes can be considered as E-branes of IIB* theory. 
Another (possibly related) problem is that the space of states of the 
corresponding open string contains negative norm states. Although 
the discussions here may clarify some of these features of 
spacelike branes,  
our main focus in this paper is on the generic properties of branes
in curved backgrounds which the spacelike branes in the plane wave 
illustrate. We thus view these branes as a useful toy example, regardless
of their physical significance, where string computations that illustrate
the features of interest are possible. 

This paper is organized as follows. In the next section 
we discuss the example of the supersymmetric $AdS_2-AdS_2$
configurations and their dual interpretation as a two monopole
solution of ${\cal N}=4$ SU(N) SYM theory. Then in section 3
we discuss spacelike branes in the plane wave. In particular,
we develop the modified lightcone gauge in section 3.1.
In section 3.2 we show that the spacelike branes discussed
are actually E-branes of IIB* theory. In the remaining 
sections we analyze in detail the open-closed duality,
the behavior of integrated amplitudes and the special
properties when the branes are at distinguished separations.

\section{Branes in AdS}

A notable example of the phenomena discussed above
is branes in $AdS \times S$ backgrounds which are separated on the sphere.
The arguments of the previous section imply that a pair of
branes located at antipodal points of the sphere have special properties.
In particular if they preserve compatible supersymmetries
the system should be supersymmetric and stable. This leads to a
number of new supersymmetric brane configurations which we will illustrate
via the specific case of $AdS_2$ branes in an $AdS_5 \times S^5$
background, although we will also discuss generalizations
at the end of this section.
One should note that since these particular branes
decouple in the Penrose limit \cite{ST} this case is not related
to branes in the plane wave.

\subsection{A supersymmetric $AdS_2$-$AdS_2$ configuration}

It was shown in \cite{ST} that a given $AdS_2$ brane located at any
point in the $S^5$ preserves half of the supersymmetries. Now
consider two such branes separated on the $S^5$. There is clearly
a distinguished configuration in which the branes are at antipodal
points on the sphere. For generic separations there is precisely
one geodesic between the branes, whilst for this configuration
there is an infinite family of geodesics since the second brane
is placed at a focusing point.
This behavior should be reflected in the spectrum of open strings
stretching between the branes: at the antipodal separations one
should get a family of zero modes for the spherical coordinates
whilst for generic separations there are no Dirichlet zero modes.

Antipodal separations are also distinguished by supersymmetry. Let
us write the $AdS_5 \times S^5$ metric as
\be ds^2 = R^2 \left (
\frac{du^2}{u^2} + u^2 (dx \cdot dx)_4 + d\q_1^2 + \sum_{k=2}^{5}
\prod_{j=1}^{k-1} \sin \q_j^2 d \q_k^2 \right ).
\ee
Following the conventions of \cite{ST}, the Killing spinors can be written as
\be
\ep = - u^{-\frac{1}{2}} \G_{4} h(\q_a) \lambda_2 +
u^{\frac{1}{2}} h(\q_a) (\lambda_1 + x \cdot \Gamma_{x}
\lambda_2),
\ee
where $\Gamma_{m}$ are tangent space gamma matrices ($4$ is the radial
direction) and $(\lambda_1,\lambda_2)$ are constant complex spinors
of negative and positive chirality respectively such that
\be
\lambda_{1} = \lambda_{1}^{R} - i \G^{0123} \lambda_1^{R}; \hsp
\lambda_2 = \lambda_{2}^{R} + i \G^{0123} \lambda_2^{R},
\ee
for real $(\lambda_1^R, \lambda_2^R)$. The function $h(\q_a)$ is
given by
\be
h(\q_a) = \exp(\half \q_1 \G^{45}) \exp( \half \q_2 \G^{56})
\exp ( \half \q_3 \G^{67}) \exp( \half \q_4 \G^{78})
\exp ( \half \q_5 \G^{89}).
\ee
The $AdS_2$ branes we discuss extend along the time and radial direction
of $AdS_5$ and they are located at a constant position
in transverse space, $(x_0^i, \theta_a), i{=}1,2,3,\ a{=}1,..,5$.
The supersymmetries preserved by an $AdS_2$ brane can be
determined via the kappa symmetry projection to satisfy
\be
\lambda_1^{R} = \eta e^{- \q_1 \G^{45}} \G^{1234} \lambda_1^{R};
\hsp \lambda_{2}^{R} = \eta e^{- \q_1 \G^{45}} \G^{1234}
\lambda_{2}^R - 2 \eta e^{- \q_1 \G^{45}} x_{0}^i \G^i \lambda_1^R,
\ee
where $\eta = \pm 1$ for a brane/anti-brane, respectively.\footnote{
Note that the worldvolume theory of a brane
is distinguished from that of an anti-brane by the
sign of the Wess-Zumino term.
Each action is invariant under kappa symmetry
but the transformations differ as
$\delta \theta = (1 - \eta \G) \kappa$.}

So as one moves a brane from the north pole of the
sphere the supersymmetries preserved by it are rotated, until at
the south pole it preserves precisely the opposite
supersymmetries: it has become an anti-brane. Therefore, a brane
(i.e. $\eta =1$) located at $\q_1 = 0$ and an anti-brane ($\eta =
-1$) at $\q_1 = \pi$, located at the same transverse positions
$x_0^i$ preserve exactly the same sixteen supersymmetries.
This should be reflected in the presence of fermion zero modes in
the open string spectrum for this configuration, and the vanishing
of exchange diagrams between these branes.

Furthermore, if these branes
are separated in their transverse positions they still preserve eight
supercharges. Branes at the same $x_{0}^i$ position preserve
eight of the ordinary supercharges and eight of the conformal
supercharges, reflecting the residual unbroken conformal invariance
$SO(2,1) \subset SO(2,4)$. Separated $AdS_2$ branes however break the
remaining conformal invariance and conformal supersymmetries.

\subsection{Dual description as a two-monopole configuration}

The dual description of the $AdS_2$ brane is as a monopole in the
gauge theory \cite{diac,Minahan, ST}, and the properties described above
correspond to known properties of monopoles in $\cN=4$ SYM theory.
The relevant fields are the vector $A_{\mu}$ and the six scalars $X^{A}$ which
transform in the $\bf{6}$ of the $SO(6)$ R symmetry. One can
define an $SO(6)$ vector of magnetic charges
\be \label{magn}
T^{A} ={1 \over v} \int_\Sigma d \Sigma^{ij} {\rm {Tr}} F_{ij} X^A,
\ee
where the integral is over a closed spatial surface that encloses
the monopole and $v^2{=}\langle X^2 \rangle$ is
the magnitude of the vev of the scalars at infinity.

Now recall the asymptotic behavior as $r \rightarrow \infty$ of
an elementary single monopole in gauge group $SU(2)$ located at
the origin in $R^3$:
\be
F^{a}_{ij} \sim \frac{g}{r^4} \ep_{ijk} x^a x^k; \hsp
X^{a A} \sim \eta \nu^{A} v \frac{x^a}{r},
\ee
where $g$ is the magnetic change,
$a$ is the gauge group index and $x^i$ parametrize
the spatial $R^3$ with $x^i x^i = r^2$. The unit vector $\nu^A$
describes the $SO(6)$ direction in which the monopole is pointing
whilst $\eta = \pm 1$ with the two signs corresponding
to monopole and anti-monopole, respectively. Fixing convenient normalizations
the mass of a single monopole $m$ is given by $m = v g$. More generally
the BPS bound may be stated as \cite{Fayet}
\be
M^2 = v^2 (T^A T^A + Q^A Q^A)
\ee
where $Q^A$ denote the electric charges (defined as in (\ref{magn}) but
with $F \to *F$).

Let us discuss more generally monopoles in $SU(N)$ gauge theory. The gauge
group is considered to be Higgsed to the maximal torus $U(1)^N$.
Recall that the monopole charges are represented by a set
of $N-1$ integers, $(m_1,...,m_{N-1})$. One can obtain
monopole solutions by embedding $SU(2)$ solutions in the $SU(N)$
theory. The details of the construction
\cite{Weinberg:1979zt} are not needed here, but we briefly review the results.
Each simple root of the $su(N)$ defines an $SU(2)$ subgroup
and the corresponding monopole solution carries a unit of magnetic
charge. These are the fundamental monopole solutions.
All other roots are associated with a superposition of fundamental
monopole solutions at the same point. Writing the root as a sum
of simple roots one obtains the constituency of the multi-monopole
solution. $su(N)$ has $N-1$ simple roots $\a^a$ and $(N-2)(N-1)/2$
positive (non-simple) roots given by
$
\sum_{a=b}^{c} \a^a, \ b  < c.
$
Any of these roots is associated with a superposition of $(c-b+1)$
elementary monopoles. In particular, there are $(N-2)$ two monopole
solutions. Explicit expressions for solutions are given
in \cite{Weinberg:1979zt}, but these will not be needed here.

So far we did not consider the effect of the global $SO(6)$ symmetry.
Consider the solution corresponding to the superposition of
one monopole and one anti-monopole pointing in a different direction
in the $SO(6)$.
Vector addition of the magnetic charges implies that
\be
|T^A| = 2 g \sin \left({\q \over 2} + {\pi \over 4} (1+\eta) \right),
\ee
where $\q$ is the angle between the two directions.
Since both the monopole and the anti-monopole have mass equal to $g$,
the BPS bound above is clearly saturated if the second
object is an anti-monopole and the angular separation is $\q = \pi$,
and in this case the total charge is two. The other possibility
is if the second object is a monopole at the same angular position.

These facts have a very simple explanation in terms of D-branes \cite{diac}.
Let us consider a configuration of $N$ D3 branes.
The worldvolume theory is a $\cN=4$ $SU(N)$ theory.
We now Higgs the theory by separating the branes in one direction, say
$X^1$,
\be
X^{1 a} = \mu_a, \qquad a=1,\cdots, N.
\ee
As discussed in \cite{diac}, the elementary monopole solutions
correspond to D1-branes stretched between consecutive D3 branes $(a,a+1)$.
D1-branes stretched between two non-consecutive D3-branes
$(a,b)$ correspond to a superposition of $b-a+1$ elementary monopoles
and carry magnetic charge $(\vec{0}_{a-1},\vec{1}_{b-a+1},\vec{0}_{n-b-1})$,
where $\vec{A}_{a}$ denotes a row vector with $a$ entries equal to $A$.
The D1 brane can be considered as a collection of D1 branes
$(a,a+1)$,..., $(b-1,b)$
with endpoints identified. Notice that a brane ending on a pair
of D-branes induces a monopole on one and an anti-monopole
in the other. The direction of the monopoles, however, in the $SO(6)$
space is also opposite and as we discussed in the previous paragraph the
total magnetic charge adds up.

Let us now consider the configuration of $N-2$ coincident branes
and two branes separated in opposite directions such that
$X^2 = v^2$ and take the near-horizon limit.
The $N-2$ coincident branes
are replaced by $AdS_5 \times S^5$, the two separated branes
were pushed to infinity and the D1 branes become the two
$AdS_2$ branes located at the antipodal points of the $S^5$
(recall that the position on the sphere of the $AdS_2$ brane is mapped
to the direction of the scalar field of the gauge theory monopole).
Thus we find direct agreement with the bulk result.

When the monopoles are separated in the $R^3$, the scale
introduced necessarily leads to the breaking of conformal symmetry
and of the conformal supersymmetries. Nonetheless the configuration
is $1/4$ BPS.

Under S duality the $AdS_2$ D-brane becomes a $(p,q)$
$AdS_2$ string, whilst the monopole is mapped into
a $(p,q)$ dyon in the gauge theory. These objects should
exhibit similar properties, which could be demonstrated
for the strings using the manifestly $SL(2,Z)$ covariant formulation
of \cite{Ceder}.

\subsection{Generalizations}

The discussion here generalizes to all cases where the background involves
spheres. A particularly interesting class of such examples are branes
on $AdS_k \times S^l \times S^m \times T^n$ for $k,l,m=2,3$
(with one or two sphere factors and $n$ such that the spacetime has $D=10$). 
These spacetimes are derived as a 
near-horizon limit of
brane intersections and are exact solutions of string theory, i.e.
there are WZW models associated with them \cite{BPS}. Branes on
$S^3$ and $AdS_3$ have been extensively analyzed in recent years,
see \cite{AdS3}
for an (incomplete) set of references. In these cases one
should be able to go beyond the supergravity approximation
and explicitly compute the relevant string amplitudes. We will not
pursue this here. Instead we will discuss a different set of examples
where the exact computation of cylinder is also possible, namely
branes in the plane wave background of IIB supergravity.

Another generalization involves spacelike branes on $AdS$ spacetimes.
Recall that (the universal cover of) global $AdS_{d+1}$ has the metric
\be
ds^2 = R^2
\left ( - \cosh^2 \rho
dt^2 + d \rho^2 + \sinh^2 \rho d \Omega_{d-1}^2 \right ), \qquad
0 \le \rho < \infty, \ -\infty < t < +\infty
\ee
Timelike geodesics are periodic with period $2 \pi$ and they
reconverge after $t=\pi$ \cite{AIS}. (The former follows from the fact
that before we take the universal cover the time coordinate
had range $[-\pi, \pi]$). It follows that the system of
spacelike branes located at $\rho=0, t=k \pi,\ k \in Z$
should exhibit special properties. In the next few sections
we analyze related branes in the plane wave background
of IIB supergravity.

\section{Spacelike branes in the plane wave}

The second example that we analyze in detail is the case 
of spacelike branes in the maximally supersymmetric plane
wave background. For our purposes it is crucial that the 
branes are spacelike and are in a Lorentzian background.
The reason is that (as we discuss in detail later) 
the Lorentzian spacetime has focal points in the 
$x^+$ direction. Wick rotating to an Euclidean section by 
$x^+ \to i x^+$, so that one would now be discussing D-instantons,
results in geodesics that are no longer periodic in $x^+$. 

The boundary 
states for the branes under consideration have been discussed previously
in \cite{billo,BGG,ST3}. Here the emphasis is on the fact that 
the worldsheet theory and the target spacetime are Lorentzian.
The standard lightcone gauge does not allow for a description
of spacelike branes in the open string channel. We thus 
develop in detail in the next subsection a modified lightcone 
gauge that allows for such a description.

The branes we discuss have imaginary couplings to the RR fields.
This can be read off from the corresponding boundary state.
So these branes are not S-branes.
They are however E-branes: they have real couplings when considered as 
branes of the IIB* plane wave. In fact we 
show in section \ref{tdualsec} that (formal) T-dualities along the lightcone 
map the Lorentzian $(+,-,m,n)$ branes of the IIB plane wave to the Euclidean
$(m,n)$ branes of the IIB* plane wave. 

\subsection{Open strings in a modified lightcone gauge}

In this section we discuss a modified bosonic lightcone gauge,
appropriate for describing certain classes of D-branes and for checking
open/closed duality. We will discuss the use of this gauge for strings in
the plane wave; the flat space case is also clearly contained in
this discussion by setting the mass parameters to zero.
The action for strings in the plane wave, with Brinkmann metric
\be \la{pw}
ds^2 = 2 dx^+ dx^- + \sum_{I=1}^{8} (-  \mu^2 (x^I)^2 (dx^+)^2 + (dx^I)^2),
\ee
and RR flux
\be
F_{+1234} = F_{+5678} = 4 \mu,
\ee
in fermionic lightcone gauge is \cite{Met}
\bea
S &=& T \int_\Sigma d^2\s \left ( -\half \sqrt{-g} g^{ab}
(2 \del_{a} x^{+} \del_{b} x^{-} - \mu^2 x_{I}^2 \del_{a} x^{+} \del_{b} x^{+}
+ \del_{a} x^{I} \del_{b} x^{I}) \right .\label{kfix} \\
&& \left . \hsp - i
\sqrt{-g} g^{ab} \del_{b} x^{+} (\bar{\q} \del_{a} \q
+ \q \del_{a} \bar{\q} + 2 i \mu \del_{a} x^{+} \bar{\q}
\Pi \q ) + i
\ep^{ab} \del_{a} x^{+} (\q \del_{b} \q +
\bar{\q} \del_{b} \bar{\q}) \right ). \nn
\eea
In this expression $g_{ab}$ is the worldsheet metric with $(\t,\s)$
the worldsheet coordinates and $\ep^{01} =1$. $(x^{+},x^{-},x^{I})$
are the bosonic coordinates of the target superspace.
$\q = \frac{1}{\sqrt{2}} (\q^1 + i \q^2)$ is a complex $SO(8)$ spinor
of positive chirality.
The $8 \times 8$ matrices $\g^{I}_{a \dot{b}}$ and its transpose
$\td{\g}^{I}_{\dot{a} b}$ are the off-diagonal components
of the $16 \times 16$ $SO(8)$ $\gamma$
matrices and couple $SO(8)$ spinors of opposite chirality. The matrix
$\Pi = \g^{1} \td{\g}^2 \g^3 \td{\g}^4$. Fixing $\a' = 1$, $T$ is
the inverse length of the string, which we choose to be
$2\pi$ ($\pi$) for a closed (open) string respectively.

The standard lightcone gauge choice $x^+ = p^+ \t$ with
conformal gauge $g_{ab} = \eta_{ab}$ leads to an action for
free massive fields. This is however not the only simplifying gauge
choice: the more general gauge choice
\be
x^{+} = x_{0}^+ + p^{+} \t + r^+ \s,
\ee
along with the conformal gauge $g_{ab} = \eta_{ab}$ also leads
to a purely quadratic action. With such a gauge choice the action
becomes
\bea
S &=& T \int d^2 \s (p^+ \pa_{\t} x^- - r^+ \pa_{\s} x^-
+ \frac{1}{2} ( (\pa_{\t} x^I)^2 - (\pa_{\s} x^I)^2 - \mu^2 a_{+} a_-
(x^I)^2) \\
&& + i (a_{-} \q^1 \pa_+ \q^1 + a_{+} \q^2 \pa_{-} \q^2 - 2 \mu
a_{-} a_{+} \q^1 \Pi \q^2), \nn
\eea
where $a_{\pm} = (p^+ \pm r^+)$. So far we have not imposed any
worldsheet periodicity or boundary conditions. For closed strings
the gauge choice will only be consistent with periodicity
when $r^+ = 0$ (unless the spacetime coordinate $x^+$ is compactified).
Thus in the closed string channel one should use the standard
lightcone gauge choice, which immediately enforces that any
boundary states at fixed $\t$ are Dirichlet in $x^+$.

In the open string channel, the general gauge choice
can be applied. However, one is usually interested in describing
D-branes with pure Neumann or pure Dirichlet boundary conditions
at fixed $\s$. For these cases, one must impose
the gauge choices $r^+ = 0$ and $p^+ = 0$ respectively.

An interesting alternative possibility would be to impose $a_{+} =
0$ or $a_{-} =0$. We will not explore this here, except for the
following comments. These cases correspond to a ``hybrid'' gauge
where the static gauge $X^0=\tau$, $X^1=\s$ (for $a_+=0$, when
$a_-=0$ we have $X^1=-\s$) is chosen for the bosons and fermionic
lightcone gauge is chosen for the fermions. This gauge however
appears somewhat singular as half of the fermions drop out
completely from the action and it may not be an admissible gauge,
see a related discussion in \cite{FST} where such a hybrid gauge for
M2 branes is considered. Furthermore, the Virasoro constraints are
more complicated than in the standard lightcone gauge.

Recall that the action in standard lightcone gauge $r^{+} = 0$
is given by
\bea
S[p^+] &=& T \int d^2\s \left ( p^{+} \del_{\t} x^{-} + \half (
(\del_{\t} x^{I})^2 - (\del_{\s} x^{I})^2 - m^2 (x^{I})^2 ) \right .
\la{sp+} \\
&& \left . + i p^+ (\q^1 \del_{+} \q^1 + \q^2 \del_{-} \q^2 - 2 m \q^1
\Pi \q^2) \right ), \nn
\eea
where $m = \mu p^+$.
The resulting action in the new gauge $p^+ = 0$ is
\bea
S[r^+] &=& T \int d^2\s \left ( - r^{+} \del_{\s} x^{-} + \half (
(\del_{\t} x^{I})^2 - (\del_{\s} x^{I})^2 +
\td{m}^2 (x^{I})^2 ) \right . \la{sr+} \\
&& \left . - i r^+ (\q^1 \del_{+} \q^1 - \q^2 \del_{-} \q^2
- 2 \td{m} \q^1 \Pi \q^2) \right ). \nn
\eea
Here $\del_{\pm} = \del_{\t} \pm \del_{\s}$ and $\td{m} = \mu r^{+}$.
Notice that the standard light cone gauge leads to a system
of 8 free bosons and 8 free fermions in a harmonic oscillator
potential, whilst the new lightcone action describes the same degrees
of freedom but in an {\it inverted} harmonic oscillator potential.
The two actions are formally
related by $m \rightarrow i \td{m}$ and $\q^2 \rightarrow - i \q^2$.
They also differ in $x^-$ boundary terms; the latter are negligible
in (\ref{sp+}) since they are total time derivatives but not in (\ref{sr+})
where they are spatial derivatives.

The field equations from (\ref{sr+}) are
\bea
(\pa_{\t}^2 - \pa_{\s}^2) x^{I} = \td{m}^2 x^{I}; \nn \\
\pa_{+} \q^1 = \td{m} \Pi \q^2; \hsp
\pa_{-} \q^2 = \td{m} \Pi \q^1,
\eea
whilst one can show that the gauge fixed Virasoro constraints are
\bea
T_{\s \t} &=& r^{+} [\del_{\t} x^{-}  + i
( \q^1 \del_{\t} \q^1 + {\q}^2
\del_{\t} {\q}^2 )] + \del_{\t} x^{I} \del_{\s} x^{I}= 0; \label{vir} \\
T_{\t\t} &=& {\cal{H}}_o =
[r^{+} \del_{\s} x^{-} + i r^+ ({\q}^1 \del_{\s} \q^1
+ \q^2 \del_{\s} {\q}^2 - 2 \td{m} \q^1 \Pi \q^2)] \nn \\
&& + \half ( (\del_{\t} x^I)^2 + (\del_{\s} x^{I})^2 -
\td{m}^2 (x^I)^2 ) = 0.
\la{H}
\eea
Note that the latter is the canonical Hamiltonian for the action (\ref{sr+}).
It is convenient to write
\be
{\hat{\cal{H}}}_{o} = \half
( (\del_{\t} x^I)^2 + (\del_{\s} x^{I})^2 - \td{m}^2 (x^I)^2 )
+ i r^+ (\q^2 \pa_{\t} \q^2 - \q^1 \pa_{\t} \q^1),
\ee
so that ${\cal{H}}_{o} = r^+ \pa_{\s} x^- + {\hat{\cal{H}}}_{o} = 0$
when one imposes the fermion field equations.

Consistent boundary conditions at $\s = 0, \pi$
following from the variational problem are pure
Neumann ($\pa_{\s} X^r = 0$) and Dirichlet ($\pa_{\t} X^{r'} = 0$)
conditions for the bosons and $\q^1 = i M \q^2$ for the fermions, where
$M$ is an orthogonal matrix which is the product of gamma matrices
$\g^{r'}$ where $r'$ are the Dirichlet directions transverse to
the lightcone. (\ref{vir}) then immediately enforces that $x^-$ is
Dirichlet, as $x^+$ manifestly also is. The physical meaning of these
boundary conditions is clear: we are quantizing open strings stretched
between two {\it spacelike} branes in Lorentzian spacetime,
separated along the lightcone as well as in spatial directions.
Note that our convention is as in \cite{Hull} that an Ep-brane has
$p$ longitudinal spacelike directions.

As discussed in $\cite{ST}$, branes in the plane wave are
naturally divided into classes depending on which directions transverse
to the lightcone they span. Here
we focus on $(m,m+2)$ branes for which $(M \Pi)^2 = -1$; in
the classification of $\cite{DP,ST2,GG,ST3,Skenderis:2003wx}$ 
these are $D_{-}$ branes.
Since we are interested in generic properties of spacelike branes which
depend on the background geometry rather than the specific class of branes
we consider, it is convenient to focus on one class of branes. 

For generic $r^+$ the mode expansions are then as follows. For the bosons,
\bea
x^{r}(\s,\t) &=& x_{0}^r \cosh (\td{m} \t) + \td{m}^{-1} p_{0}^{r}
\sinh (\td{m}\t) + i \sum_{n \neq 0} \w_{n}^{-1} \a^r_{n} e^{-i \w_n \t}
\cos (n \s); \label{neu} \\
x^{r'}(\s,\t) &=& x_{1}^{r'} \cos (\td{m} \s) + (x_2^{r'} {\rm{cosec}}
( \td{m} \pi) - x_{1}^{r'} \cot (\td{m} \pi)) \sin (\td{m} \s) \nn \\
&& \hsp \hsp
+ \sum_{n \neq 0} \w_{n}^{-1} \a_{n}^{r'} e^{-i \w_{n} \t} \sin (n\s).
\label{dir}
\eea
In these expressions, $\w_n = {\rm{sgn}}(n) \sqrt{n^2 - \td{m}^2}$.
When $\td{m}^2 < 1$, i.e. when the mass scale set by the flux is
smaller than the string mass, all stringy modes have real frequencies.
In this regime, the upside down harmonic oscillator potential
affects mostly the zero modes: the zero modes exhibit an exponential
behavior but all stringy modes are oscillatory, as in flat space.
When $\td{m}^2 > 1$ stringy modes with $n^2 < \td{m}^2$
exhibit exponential behavior. This case can be included in
the generic analysis with $\w_n \rightarrow i {\rm{sgn}}(n)
\sqrt{\td{m}^2 - n^2}$.
The generic analysis breaks down when $\td{m}$ is integral.
In this case one stringy mode becomes massless.
We will discuss in detail the physical significance of these values later.
For concreteness, we consider $\td{m}^2 < 1$ in the generic analysis.

The commutation relations are
\be
[x_{0}^r, p_0^{s}] = i \d^{rs}; \hsp
[\bar{a}_{0}^r, a_0^s] = \d^{rs}; \hsp
[a_n^{I}, a_l^{J}] = {\rm{sgn}}(n) \d_{n+l} \d^{IJ},
\ee
where one defines 
(as noted, $\td{m}$ is
assumed non-integral, so $\w_n \neq 0$).
\be \label{defa}
a_{0}^r = e^{-i \pi/4} \frac{1}{\sqrt{2\td{m}}} (p_0^{r} - \td{m} x_0^r), \hsp
\bar{a}_0^r = e^{-i \pi/4} \frac{1}{\sqrt{2\td{m}}}
(p_0^r + \td{m} x_0^r), \hsp
a_n^{I} = \sqrt{\frac{1}{\left | \w_n \right |} } \a_n^{I}.
\ee
The phases in the definition of $a_0$ and $\bar{a}_0$ are needed in order
for their commutator to be real.
The fermion mode expansions are the following
\bea
\sqrt{r^+} \q^1 &=& \q_0 \cosh (\td{m} \t) + \tilde{\q}_{0} \sinh (\td{m} \t)
+ \sum_{n \neq 0} c_n^2 \left ( d_{n} (d_n - M \Pi) \q_n \phi_n
+ \td{\q}_{n} \td{\f}_{n} \right ) ; \label{fer} \\
\sqrt{r^+} \q^2 &=& \Pi \td{\q}_0 \cosh (\td{m} \t)  + \Pi {\q}_{0}
\sinh (\td{m}\t)
+ \sum_{n \neq 0} c_n^2 \left ( - i d_n \Pi \td{\q}_n \td{\f}_n
+ i (M^t + d_n \Pi) {\q}_{n} \f_{n} \right ), \nn
\eea
where
\bea
\f_{n} = e^{-i ( \w_n \t + n \s)}, \hsp
\td{\f}_n = e^{-i ( \w_n \t - n \s)}; \nn \\
d_n = \frac{1}{\td{m}} (\w_n - n), \hsp
c_n = \frac{1}{\sqrt{1+d_n^2}}.
\eea
Imposing the boundary conditions one gets
$\td{\q_0} = i M \Pi \q_0$ and $\td{\q}_n = -( 1 + d_n M \Pi) {\q}_n$
whilst the anticommutators are given by
\be
\left \{ \q_0^a, \q_0^b \right \} = \qu \d^{ab} ; \hsp
\left \{ \q_{n}^a, \q_m^b \right \} = \qu \d^{ab} \d_{n+m}.
\ee

Because of the fermion boundary condition\footnote{$\q|$ indicates
evaluation of $\q$ at $\s=0,\pi$.}, $\q^1|=i M \q^2|$,
the fermions cannot be taken to be real. This reflects
the fact that the brane is spacelike. However, one can still impose
a modified reality condition: 
\be
(\q^1_a)^{*} = A_a^b \q^1_b, \qquad (\q^2_a)^{*} = B_a^b \q^2_b
\ee
where $A$ and $B$ are $8 \times 8$ real matrices, i.e. the complex
conjugate of the spinor is not the spinor itself but a real rotation
of it. The matrices $A$ and $B$  should be equal to a sum
of even powers of gamma matrices so that they do not
change the chirality of the spinor.
Compatibility with the field equations, boundary conditions
and the fact that $*$ should be an involution yields (after some
manipulations),
\be \label{Rcod}
A^2 =1, \qquad  B = \Pi A \Pi, \qquad \{M \Pi, A\}=0.
\ee
One can obtain the action of the new reality condition on the
modes as
\be \label{reality}
(\q_0)^{*} = A \q_0, \qquad (\q_n)^{*} = A \q_{-n}.
\ee
The bosonic oscillators satisfy the standard reality conditions,
i.e. $p_0^r$ and $x_0^r$ are real and $a_n^* = a_{-n}$, but
because of the phase factors in (\ref{defa}) $a_0$ and $\bar{a}_0$ 
satisfy unconventional reality conditions, 
\be
a_0^* = -i a_0, \qquad \bar{a}_0^* = -i \bar{a}_0.
\ee

The solution to (\ref{Rcod}) differs depending on whether
$M^2=1$ or $M^2=-1$. The former case corresponds to
E4 branes and in this case
\be
M^2 =1: \qquad A=M, \quad B=-M.
\ee
Thus in this case the reality condition is consistent
with the isometry group preserved by the E4 brane.
The case $M^2=-1$ is relevant for E2 and E6 branes.
It is easy to see that (\ref{Rcod}) admits a solution in all cases,
but one has to select two directions, one in each $SO(4)$.
Consider for example a $(2,0)$ brane extending along
the 1 and 2 directions. Then one has to select two directions,
one in each $SO(4)$ and both transverse to the brane. For instance
one may select the directions 4 and 8, so that $A=\g^{3567}$ (and
consequently $B=-\g^{1248}$) is a solution. Thus in this
case a choice of a reality condition breaks further the
(bosonic) isometry group.

With an appropriate choice of basis, 
$A = \s^3 \otimes 1_4,\ M\Pi = i \s^2 \otimes 1_4$, 
where $\s^i$ are Pauli matrices 
and $1_4$ is a $4 \times 4$ matrix. Let us also define
\be
\q=\left(
\begin{array}{c}
\phi \\
i \l
\end{array}
\right)
\ee
where $\phi$ and $\l$ are 4-component spinors.
The reality condition (\ref{reality}) implies
\be
\phi_n^*= \phi_{-n}, \qquad \l_n^*=\l_{-n}.
\ee
In terms of this decomposition 
the anticommutation relations read
\be
\{\phi_0, \phi_0\} = {1 \over 4}, \qquad \{\l_0, \l_0\} = - {1 \over 4}, \qquad
\{\phi_n, \phi_{-n}\} = {1 \over 4}, \qquad 
\{\l_n, \l_{-n}\} = - {1 \over 4},
\ee
where we suppress the spinor indices. Thus the $\l$-modes have ghost-like 
anticommutation relations and the state space has indefinite metric.
 
The generator of $\s$-translations is given by the conserved charge
$
\hat{H}_{o} = \frac{1}{\pi} \int^{\pi}_{0} d \s {\hat{\cal{H}}}_o
$
with mode expansion 
\bea
\hat{H} _{o} &=& h_D + h_{0} + h_{N}; \\
h_{D} &=& \frac{\td{m}}{2 \pi \sin(\td{m} \pi)} \sum_{r'}
( \cos ( \td{m} \pi) (( x_1^{r'})^2 + (x_2^{r'})^2)
- 2 x_1^{r'} x_2^{r'}); \nn \\
h_0 &=& 
\half \sum_{r=1}^{p}((p_0^r)^2 - \td{m}^2 (x_0^r)^2) 
+ 2 \td{m} \q_0 M \Pi \q_0 
=\half \sum_{r=1}^{p}((p_0^r)^2 - \td{m}^2 (x_0^r)^2) 
+ 4 i \td{m} \phi_0 \l_0; 
\nn \\
h_N &=& \sum_{n > 0} ( \w_n a^{I}_{-n} a^{I}_n + 4 \w_n \q_{n} \q_{-n}) 
=\sum_{n > 0} ( \w_n a^{I}_{n}{}^\dagger a^{I}_n 
+ 4 \w_n (\f_{n} \f_n^\dagger - \l_n \l_n^\dagger)). \nn
\eea
which is clearly hermitian but not positive definite. 

In proceeding to quantize the system one is faced with the problem that the 
Hamiltonian is unbounded from below and half of the fermionic modes
satisfy ghostlike anticommutation relations. As we discuss in the 
next section these branes are T-dual along the lightcone directions
to standard timelike branes. T-duality suggests that the 
appropriate quantization is the ``analytic'' continuation of the 
quantization of the string in the standard harmonic oscillator 
potential. One could argue that the problems 
we encounter here are associated with the fact that 
one of the T-dualities is timelike. Proceeding in this way we
define the vacuum by
\be
\bar{a}_0 \vac = a_n^{I} \vac = 0, \qquad
\q_{-n} \vac = 0 \qquad
(1 - i M \Pi) \q_0 \vac = \q_0^- \vac = 0.
\ee
where we define $\q_0^{\pm} = 2 (\phi_0 \pm \l_0)$. These modes 
satisfy the anticommutation relations, 
\be
\{\q_0^+,\q_0^-\} =1, \qquad \{\q_0^{\pm}, \q_0^{\pm}\}=0.
\ee

In summary, we consider $\bar{a}_0, \q_0^-, 
a_n, \q_{-n}$ as annihilation operators 
and  $a_0, \q_0^+$,$ a_{-n}, \q_{n}$ with negative $n$ as creation
operators. $|0\rangle$ is annihilated by all annihilation
operators and $\langle 0|$ by all creation operators. 
We now build the Fock space by acting on $|0\rangle$ 
with the creation operators (or on $\langle 0|$ by
annihilation operators). Notice that the bar and ket 
states are not related by conjugation (because of the fermion
zero modes), but there is a natural inner product. 

The spectrum constructed this way is identical to the spectrum 
of Lorentzian $(+,-,m,n)$ \cite{ST3} but the eigenvalues are related by 
$m \to i \td{m}$. In particular, the states generated by the zero modes 
have imaginary eigenvalues w.r.t. $\hat{H}_0$ {}\footnote{
An alternative quantization that would avoid imaginary eigenvalues
for the states built from the bosonic zero modes
has recently been discussed in \cite{Pioline:2003bs}:
the inverted harmonic oscillator admits a continuous spectrum 
of delta-function normalizable scattering states, and one could consider those
instead of the discrete states discussed here. These states lead to the 
same one loop amplitude as the discrete states.
For our purposes one would need to extend the discussion of 
\cite{Pioline:2003bs} to include the states build from the 
fermionic zero modes.}. This is not in contradiction 
with the fact that  $\hat{H}_0$ is formally hermitian, as the state space 
has indefinite metric.

\subsection{T-duality and relation 
with E-branes of type IIB* theory} \label{tdualsec}

In flat space, one can also view the spacelike branes as being related
by formal T-duality in the $(x^+,x^-)$ directions to the usual
Lorentzian branes. Under such T-duality, in one timelike and one
spacelike direction, the type IIB theory is mapped to the type IIB*
theory \cite{Hull} in which the RR fields have opposite
sign kinetic terms to usual. Thus the boundary states describe
E-branes in the type IIB* theory. Notice that this argument applies
irrespectively of whether one is using the lightcone GS or the RNS 
description. A detailed comparison between the lightcone GS and 
RNS descriptions of branes satisfying Dirichlet conditions along
the time direction for the case of $p=-1$ can be found in \cite{Green:iw},
and it seems likely that the conclusions of this paper
extend to all other $p$. Provided that this is the case, some of the 
boundary states that have been proposed to 
describe S-branes, such as the RNS boundary state given in section 4.1 of
\cite{Gutperle:2002ai} which is pure Dirichlet in the time direction, 
contain imaginary couplings to the RR-fields
and as such should be associated with E-branes and not S-branes.

In the plane wave the same interpretation holds. To see this
we first work out how the T dualities act on the plane wave
background. We define new coordinates
$
x^{
\pm} = {1 \over \sqrt{2}} (\pm t + u)
$
and T-dualize on $(u,t)$ using standard T
duality rules \cite{Tdual}. This (formal) procedure results in the
following solution for the T dual background:
\bea
ds^2 &=& 2 d\tilde{x}^+d\tilde{x}^-
+ \sum_{I=1}^{8} ( \mu^2 (x^I)^2 (d\tilde{x}^+)^2 + (dx^I)^2); \\
\hat{F}_{+1234} &=& \hat{F}_{+5678} = 4 \mu. \nn
\eea
The dual lightcone coordinates are related to the dual $(\tilde{t},\tilde{u})$
coordinates as
$
\tilde{x}^{\pm} = {1 \over \sqrt{2}} (\mp \tilde{t} + \tilde{u}).
$
There is also an imaginary shift by $\pi/2$ of the dilaton,
but this just reflects the fact that the NSNS and RR fields
in the IIB* theory have opposite signs for their kinetic terms.
The difference compared to the usual plane wave (\ref{pw})
is that the sign of the $g_{++}$ term is reversed.
This means that the T-dual background
is a real solution of IIB* theory rather than IIB theory,
as expected since we T-dualized along the lightcone. Let us call
this background the IIB* plane wave. Above we denote
by ${\hat{F}}$ a RR field in the IIB* theory, to distinguish it
from the RR fields in ordinary type IIB.
The IIB* plane wave is clearly related to the ordinary IIB
plane wave by the analytic continuation $\mu \rightarrow i \mu$
which acts on the RR field strength as
$F \rightarrow i F \equiv \hat{F}$.

Now consider the action of these T-dualities on branes. A Lorentzian
$(+,-,m,n)$ brane in the IIB plane wave is mapped to a $(m,n)$ brane
in the IIB* plane wave. The spectrum and amplitudes of the latter
will be related to those for a $(m,n)$ brane in the IIB plane
wave via analytic continuation $\mu \rightarrow i \mu$. This explains
the relationship between the spectra of Lorentzian branes
and Euclidean branes in the IIB plane wave.

\subsection{Open/closed duality}

Since one can describe spacelike branes in the closed
channel with the usual lightcone gauge and in the open
channel by the modified lightcone gauge choice, it is straightforward
to check the Cardy consistency condition. Most of the ingredients
required are given in the discussions of \cite{BGG,GG}, in
particular, the modular transformations of the relevant
mass deformed modular functions.

However, these papers leave
open the issue of whether the cylinder condition is satisfied
for branes displaced from the origin in transverse space.
Indeed, it has been suggested that the lack of dynamical
supersymmetry for such displaced branes could lead to a violation
of the Cardy condition. Here we show that
the amplitudes for displaced branes do satisfy the cylinder condition
and  that such displaced spacelike branes are annihilated by
the same number of supercharges as the branes at the origin.
Moreover, our discussion of the computation of the cylinder
amplitude in the modified open string gauge involves non-trivial new features,
namely the use of the canonical Hamiltonian rather than the non-conserved
lightcone Hamiltonian.

\subsubsection{Closed string channel}

The relevant features of the closed string mode expansions
are reviewed in the appendix.
The gluing conditions (for a boundary on the worldsheet at $\t =0$)
are
\bea
N : p_{0}^{r} \bol = 0; \hsp (\a^{1r}_{n} + \a^{2r}_{-n})
\bol = 0; \\
D : x_{0}^{r'} \bol = x^{r'} \bol; \hsp
(\a^{1r'}_{n} - \a^{2r'}_{-n}) \bol = 0, \nn \\
(\q_{0}^1 + i \eta M \q_{0}^2) \bol = 0; \hsp
(\q_{-n}^{1} + i \eta M \q^2_n) \bol = 0. \nn
\eea
where $N$ and $D$ are Neumann and Dirichlet directions respectively.
Note that $x^{r'}$ is the eigenvalue of the operator $x_0^{r'}$ and
recall that in light cone gauge $x^+$ and $x^-$ necessarily satisfy
Dirichlet boundary conditions in the closed string channel. Here
$M_{ab} = (\g^{r'_{1}} \td{\g}^{r'_{2}} .. \td{\g}^{r'_{8-p}})_{ab}$
is the product of the gamma matrices over the Dirichlet directions
and $\eta = \pm 1$ describes brane and anti-brane respectively.
The boundary state at general $x^+$ can be obtained by acting
with the time evolution operator $e^{- i H x^+/p^+}$ where $H$
is the closed string lightcone Hamiltonian given by\footnote{Note
that the definitions of conserved charges in terms of worldsheet
fields are as given in appendix B of \cite{ST3}. Our conventions
differ from those of \cite{ST3} in that we use $SO(8)$ rather
than $SO(9,1)$ spinors; the oscillators have also been rescaled
for convenience, compare appendix C of \cite{ST3} with the appendix A
here.}  
\be \la{cham}
H = \frac{1}{2} (p_0^2 + m^2 x_{0}^2) + i m
\q_0^1 \Pi \q_{0}^2 + \sum_{\ca = 1,2} \sum_{n > 0} (
\a_{-n}^{\ca I} \a_n^{\ca I} + \w_n \q_{-n}^{\ca} \q^{\ca}_n).
\ee
The closed string is invariant under 16 kinematical $Q^{+1,2}$
and 16 dynamical supersymmetries $Q^{-1,2}$.
Let us define complex combinations as follows
\bea
&& Q^+ = Q^{+1} + i \eta M Q^{+2}, \nn \\
&& Q^- = Q^{-1} + i \eta M Q^{-2} - \half i \mu \sqrt{p^+}
\sum_{r'} x^{r'} \g^{r'} M \Pi (Q^{+1} - i \eta M Q^{+2}).
\eea
As discussed in \cite{ST3}, the displaced spacelike brane
is annihilated by the supercharges
\be
Q^+ \bol = 0; \qquad  Q^- \bol = 0. \nn
\ee
i.e., the boundary state is Grassmann analytic (it is annihilated
by the $Q^{\pm}$, but not the complex conjugates).
In other words, the supercharges preserved by the boundary state
form a 16 dimensional subspace of the complex space spanned
by $Q^{\pm}, \bar{Q}^{\pm}$. The unusual reality conditions of
the preserved supercharges are related to the fact that
the corresponding worldvolume theory is spacelike.

The explicit solution for the boundary state is \cite{billo,BGG}
\bea \la{bstat}
\bol &=& {\cal{N}} \exp \left ( \sum_{n=1}^{\infty}
(\w_n^{-1} M_{IJ} \a^{I1}_{-n} \a^{J2}_{-n} - i \eta M \q^1_{-n}
\q^2_{-n}) \right ) \bo1 \\
\bo1 &=& (M_{IJ} \left | I \right > \left | J \right >
+ i \eta M_{\dot{a} \dot{b}} | \dot{a} \rangle
| \dot{b} \rangle)
e^{-\half \sum_r {a_0^r a_0^r} + \half
\sum_{r'} \left(a_0^{r'} - i \sqrt{2 m} x^{r'} \right)^2} \vac,\nn
\eea
where $M_{IJ}$ is a matrix with diagonal entries of $-1$ and $1$
for Neumann and Dirichlet directions respectively. The matrix
$M_{\dot{a} \dot{b}} = (\tilde{\g}^{r'_{1}} \g^{r'_{2}} ...
\g^{r'_{8-p}})_{\dot{a} \dot{b}}$ is the product of gamma matrices in
the Dirichlet directions. $ {\cal{N}}$ is an overall normalization,
to be fixed by factorizing the annulus computed in the open string
channel. Note that the action of the fermion zero modes
(or equivalently the kinematical supercharges) on these states
is
\be
\sqrt{2} \q_0^a | I \rangle = \g^{I}_{a \dot{a}} |
\dot{a} \rangle , \hsp
\sqrt{2} \q_0^a | \dot{a} \rangle = \td{\g}^{I}_{\dot{a} a}| I \rangle.
\ee
In tensor products such as $ | I \rangle | J \rangle $
$\q_{0}^1$ and $\q_0^2$ act on the first and second states respectively.

This representation of the fermion
zero mode part of the ground state is particularly useful
for determining supergravity field sources.
Consider first $M_{IJ} | I \rangle | J \rangle$: one decomposes
$M_{IJ}$ into $SO(8)$ representations $8 \otimes 8 = 35 + 28 + 1$.
The $35$ is the symmetric traceless part, corresponding to the
transverse graviton $h_{IJ}$; the $28$ is the antisymmetric part,
corresponding to the transverse 2-form $b_{IJ}$, and the singlet
is the dilaton $\phi$. As usual, one can choose lightcone gauge for
the supergravity fluctuations, $\psi_{- M \cdots} = 0$, and the non-dynamical
modes $\psi_{+ M \cdots}$ are determined in terms of these transverse
modes.
The RR part of the boundary state $M_{\dot{a} \dot{b} }
| \dot{a} \rangle | \dot{b} \rangle$ can also be decomposed as
\be
M_{\dot{a} \dot{b} } = \frac{1}{8} \d_{\dot{a} \dot{b} }
{\rm{tr}}(M) + \frac{1}{16} \g^{IJ}_{\dot{a} \dot{b}}
{\rm{tr}} (\g^{IJ} M)
+ \frac{1}{384} \g_{\dot{a} \dot{b}}^{IJKL} {\rm{tr}} (\g^{IJKL} M),
\ee
defining the couplings
to the RR scalar $\chi$, two-form $c^R_{IJ}$ and four-form
$c^R_{IJKL}$ respectively. Again the non-propagating components $\psi_{+ M
\cdots}$ are determined by the transverse modes.

This discussion follows that of \cite{Green:1996um}
for the supergravity sources of boundary states in lightcone
gauge in flat space.  There is an important difference in the plane wave,
however: the states $ | I \rangle | J \rangle$
and $| \dot{a} \rangle | \dot{b} \rangle $ are not eigenstates of
the Hamiltonian. To describe the lightcone time evolution of the
boundary state it is convenient to write the boundary state instead
in terms of such eigenstates, constructed in \cite{mt}. To do so one defines
complex combinations of fermion zero modes
\be \la{vac1}
\q_R = {1 \over 2 \sqrt{2}} (1 + \Pi) (\q_0^1 + i \q_0^2); \hsp
\q_L = {1 \over 2 \sqrt{2}} (1 - \Pi) (\q_0^1 + i \q_0^2),
\ee
and chooses the closed string vacuum to be such that
$\bar{\q}_{L} \vac = \q_R \vac = 0$. Then the boundary
state for branes such that $M^2 = -1$, i.e.
the $(2,0)$ and $(4,2)$ branes, is
\be \la{vac2}
\exp ( - \half \eta M_{a b}
\q_{L}^a \q_L^b + \half \eta M_{ab}
\bar{\q}_R^a \bar{\q}_R^b) \vac,
\ee
whilst an analogous expression holds for the $(1,3)$ branes
for which $M^2 = 1$; we will not need the explicit expression here.
Expanding the exponential, one can then infer the supergravity sources
by comparison with the tables given in \cite{mt}.

From the explicit form
of the boundary state (\ref{bstat}) it is immediately apparent that these
branes source purely imaginary RR fields, which is to
be expected since the branes are spacelike. Note also that
the boundary states at $x^+ = 0$ source only the graviton,
dilaton and (imaginary) RR $p$-form potential, as in
flat space \cite{DR}. Boundary states at general $x^+$
however source different supergravity fields and are not
pure position/momentum eigenstates. We will discuss later
the lightcone time evolution of the branes.

\bigskip

The cylinder amplitude between separated pairs of branes is given
by
\be
\cA (X^+, X^-, x_1, x_2) = \left \langle x_1^+, x_1^-, x_1 |
\Delta | x_2^+, x_2^-, x_2 \right \rangle,
\ee
where $\Delta$ is the closed string propagator and $(x_1,x_2)$
are the transverse positions of the branes. The branes are
also separated in the lightcone directions so that
$X^{\pm} = (x_2^{\pm} - x_1^{\pm})$. Fourier transforming along
the lightcone one gets
\bea
\cA (X^+, X^-, x_1, x_2)
&=& \frac{1}{2 \pi i} \int dp^+ dp^- e^{i p^+ X^- + i p^- X^+}
\langle -p^-, -p^+, x_1 | \frac{1}{p^+ p^- + H} | p^-, p^+, x_2
\rangle; \nn \\
&=& \int^{\infty}_{0} dp^+ e^{i p^+ X^-}
\langle -p^+, x_1 | e^{- \frac{i H X^+}{p^+} } | p^+, x_2 \rangle,
\eea
where $H$ is the lightcone Hamiltonian given in
(\ref{cham}). A suitable regularization prescription is implicit
in these expressions; we will discuss in the next section the computation
of the integrated amplitudes. One can rewrite this amplitude
as an integration over a cylinder parameter $t$
(with $X^+ = \pi p^+ t$, the $\pi$ normalization being included for later
convenience) so that
\be
\cA (X^+, X^-, x_1, x_2) = \int^{\infty}_{0} \frac{dt}{t}
e^{ i\frac{X^+ X^-}{\pi t}} \td{\cA} (t,x_1,x_2),
\ee
where
\be
\td{\cA} (t,x_1,x_2) = \left \langle -p^+, x_1 | e^{- i \pi H t}
| p^+, x_2 \right \rangle.
\ee
This amplitude is the same as that given in \cite{BGG}, except that
here we allow for non-zero Dirichlet positions.
Thus one may immediately write down the amplitude as
\be \
\td{\cA} (t,x_1,x_2) = {\cal N}_1 {\cal N}_2
{\cA}_D (1 - q^m)^{\frac{p-8}{2}}
\left ( \frac{ f_1^m(q)}{ f_1^m(q)} \right )^8,
\ee
where ${\cA}_{D}$ is the part that depends on the Dirichlet zero
modes (i.e. the position of the brane),
$q = e^{- 2 \pi i t}$ and the modular function is \cite{BGG}
\be
f_1^{m}(q) = q^{-\Delta_m} (1 - q^m)^{\frac{1}{2}}
\prod_{n=1}^{\infty} (1 - q^{\sqrt{m^2 + n^2}})
\ee
with $\Delta_m$ the Casimir energy of a boson of mass $m$ on a cylinder
with periodic boundary conditions, whose integral representation
is given in \cite{BGG}.

The part of the amplitude that depends on the Dirichlet zero mode
part is given by
\be {\cA}_{D} = \langle 0 |
e^{\frac{1}{2} \left(\bar{a}_0 + i \sqrt{2 m} x_1\right)^2}
e^{\ln(z) a_{0} \bar{a}_0} e^{\frac{1}{2} \left(a_0 - i \sqrt{2 m}
x_2\right)^2} | 0 \rangle,
\ee
where for notational simplicity
the $r'$ indices are suppressed and $z = q^{\frac{1}{2} m}$. Up to normalization, this
is the quantum mechanical amplitude $\langle x_1 | \exp i H t |
x_2 \rangle$, where $H$ is the Hamiltonian of the harmonic
oscillator. The result is well known but we present an elementary
evaluation of this amplitude in appendix B; the result is
\be
\cA_D = \frac{1}{(1-z^2)^{\hp}} e^{ - \frac{m}{1-z^2}
\left(x_1^2 + x_2^2 - 2 z x_1 x_2\right)}.
\ee
Putting this result
for the Dirichlet zero modes together with the rest of the
amplitude one gets
\be \la{ca} \cA (X^+, X^-, x_1, x_2) =
\int^{\infty}_{0} \frac{dt}{t} e^{ i\frac{X^+ X^-}{\pi t}} {\cal
N}_1 {\cal N}_2 e^{ - \frac{m}{1-z^2} (x_1^2 + x_2^2 - 2 z x_1
x_2)} \left ( \frac{f_1^m(q)}{f_1^m(q)} \right )^8. \la{clos}
 \ee

\subsubsection{Open string channel}

The one loop amplitude for the open strings in the modified
lightcone gauge is given by
\be
Z = \int_{0}^{\infty}
\frac{ds}{s} {\rm{Tr}} ( (-)^{F} e^{2 i \pi H_{o} s}),
\ee
where $H_{o}$ is the open string canonical Hamiltonian
for which one now relaxes the constraint $H_{o}=0$. 
Since the Hamiltonian generates
worldsheet time evolution, it is clearly the correct generator
to describe a loop of open strings.

One needs, however, to be careful about the sign in the exponent.
Since we are working in Lorentzian signature comparison of the
cylinder amplitudes between open and closed channels is subtle.
When one carries out an S-transformation which exchanges the sides
of a Lorentzian cylinder, this also changes the overall signature.
Thus we will need to compare a $(-1,1)$ signature cylinder in the
closed channel with a $(1,-1)$ signature cylinder in the open
channel. Our previous discussions used $(-1,1)$ signature in the
open channel and the effect of the signature change is to change
the overall sign in the Hamiltonian, $H_{o} \rightarrow - H_{o}$,
which can be seen by double analytic continuation in $\t$ and $\s$.
This explains the plus sign in the exponent above.
It is then convenient to rewrite the amplitude in the equivalent
form
\be
Z = \int_{0}^{- \infty} \frac{ds}{s} {\rm{Tr}} ( (-)^{F}
e^{- 2 i \pi H_{o} s}),
\ee
i.e. changing the overall sign of $s$.

Evaluating this amplitude one finds
\be \la{oa}
Z = \int_{0}^{- \infty} \frac{ds}{s} e^{- i \frac{\sigma^2_{10-p}}{\pi} s}
(2 \sinh (\mu X^+ s))^{4-p} \left ( \frac{ f_{1}^{\td{m}} (\td{q})}
{f^{\td{m}}_1(\td{q})} \right )^8 \la{onel}
\ee
where now
\be
f_1^{\td{m}}(\td{q}) = \td{q}^{- \Delta_{\td{m}}}
(1 - \td{q}^{i \td{m}})^{\frac{1}{2}}
\prod_{n=1}^{\infty} (1 - \td{q}^{\sqrt{n^2 - \td{m}^2}}),
\ee
with $\td{q} = e^{-2 i \pi s}$. Also $\sigma_{10-p}^2$ is
given by
\be
\sigma^2_{10-p} = 2 X^+ X^- +
\frac{\mu X^+ }{\sin( \mu X^+ )} \sum_{r' = p+1}^{8}
\left ( \cos ( \mu X^+) (( x_1^{r'})^2 + (x_2^{r'})^2)
- 2 x_1^{r'} x_2^{r'} \right ). \la{sig}
\ee
This
is the geodesic distance between two points in the plane wave \cite{Mathur},
separated by $X^{\pm}$ in the lightcone directions
and at $(x^r = 0, x_1^{r'})$ and $(x^{r} = 0, x_2^{r'})$ respectively in
the transverse directions.

The integrand obtained in (\ref{onel}) derives from the following
structure of the open string spectra. Whilst the number of bosonic
states matches the number of fermionic states at every $\hat{H}_o$
eigenvalue for stringy states, there is a mismatch between bosons
and fermions for zero modes. For $p=2$ there is a mismatch at the first
three levels of the spectrum, $\hat{H}_0 = - i \td{m}, 0,
i \td{m}$, giving rise to the
$\sinh^2$ factor. For $p=4$ it is only the vacuum state which is
unpaired, giving a factor of one,
whilst for $p=6$ there is a mismatch at every level in
the zero mode spectrum, giving the $1/\sinh^2$ factor. A detailed discussion
of the spectra in the related case of Lorentzian $D_-$ branes is given in \cite{ST3}.

\bigskip

With the conventions and normalizations used here, in the
the open string channel we have a cylinder of length $\pi$ and of
circumference $2 \pi s$ whilst in the closed string channel
the circumference is $2 \pi$ and the length is $\pi t$.
Under the S transformation $s \rightarrow - 1/t$ and
one should in addition perform a conformal transformation
so that the length and circumference of the cylinder are the
same as before. This implies that the mass parameters are
related as $\td{m}=i m t$. This can be seen as follows.
Before fixing the lightcone gauge the sigma model
was (classically) conformally invariant. In the lightcone gauge
$g_{++} \sim m^2 (x^I)^2$ and the standard conformal transformation
of $g_{++}$ implies that $m$ should transform.
As discussed above, one needs to take into account
the signature change under the S transformation, and thus the
amplitudes (\ref{clos}) and (\ref{onel}) should agree when
$s = - 1/t$ and $\td{m} = i m t$.

That the amplitudes do agree follows from the Lorentzian
S modular transformation for the mass deformed modular functions:
\be
f_1^m(e^{-2 \pi i t}) = f_{1}^{im t} (e^{\frac{2 \pi i}{t}}),
\ee
which can be derived via analytic continuation of
the proof for the Euclidean transformation given in Appendix A
of \cite{BGG}. Note that the $m \rightarrow 0$ limit of this
identity gives
\be
f_1(e^{\frac{2 \pi i}{t}}) = (i t)^{\frac{1}{2}} f_1 (e^{-2 \pi i t}),
\ee
which is the correct Lorentzian transformation property of the usual
modular function.

The cylinder amplitudes then agree provided that the boundary state
normalization is
\be
{\cal N} = (2 \sinh (\pi m))^{\frac{1}{2}(4 - p)} e^{\frac{1}{2} m \sum_{r'}
(x^{r'})^2},
\ee
and we relate the open and closed string
positions by $x_{\rm{closed}} = \sqrt{2} x_{\rm{open}}$.
The latter identification follows from the overall normalizations of
the (gauge fixed) open and closed string actions. The
$x^{r'}$ dependent normalization reflects the fact that
\be
| x^{r'} \rangle = e^{ \half m (x^{r'})^2} e^{\half (a_0^{r'} -i \sqrt{2 m}
x^{r'})^2} \vac
\ee
are the normalized position eigenstates satisfying
$\langle x_1^{r'} | x_2^{r'} \rangle = \sqrt{\pi} \d(x_1^{r'} - x_2^{r'})$.
Note that a non-trivial consistency check on these normalizations is
provided by the agreement between the amplitudes for general $(x_1,x_2)$
and for different pairs of (anti)-branes Ep-Eq ($p \neq q$), but we shall not
present the details here as similar computations for branes at the origin
were reported in \cite{BGG}.

\subsection{Behavior of integrated amplitudes}

The cylinder amplitudes vanish for cylinders which end on the same
brane; this follows from the presence of fermion zero modes.
Thus the first correction to the self energy of the brane vanishes,
presumably along with all higher corrections. The overlap between
parallel branes at the same lightcone position but
separated in the transverse directions also vanishes.

However, $D_-$ branes of the same
type but separated along the lightcone
are not annihilated by the same combinations of supercharges.
In the plane wave, the kinematical charges $Q^+$, which are represented
in terms of the fermion zero modes, do not commute with the lightcone
Hamiltonian. Thus there is non-trivial behavior for the cylinder
amplitudes between $D_-$ branes separated along the lightcone
which we now discuss. There are three cases to consider, corresponding
to $(2,0)$ E2 branes; $(3,1)$ E4 branes and $(4,2)$ E6 branes.

Note that $D_{+}$ branes, i.e. $(m,n)$ branes for which $n \neq (m \pm 2)$,
are annihilated by combinations of kinematical supercharges which
commute with the Hamiltonian \cite{ST3, GG}. This implies that the
cylinder amplitudes for such branes are zero regardless of the
brane separations. It is for this reason that we focus on $D_
-$-branes
which better illustrate the generic behavior of branes in curved backgrounds 
of interest here.  

Throughout this section we focus on the behavior of
amplitudes for generic brane separations. In the next
section we identify and discuss the physical interpretation of
distinguished separations for which the amplitudes take special values.

\subsection*{E2-branes}

In this case the cylinder amplitude is
\be
Z = \int^{\infty}_{0} \frac{ds}{s} e^{i \frac{\sigma^2_8}{\pi} s}
(2 \sinh (\mu X^+ s))^2.
\ee
The integral can be
computed by analytically continuing $x^+ \rightarrow x_{E}^+ = i x^+$;
the integral is then convergent provided that $x^-$ is positive.
Evaluating the integral under these conditions and analytically
continuing the answer to real $x^+$ and general values of $x^-$
one obtains
\be
Z = - 4 \ln (1 + \left ( \frac{2 \pi \mu X^+}{\sigma^2_8}
\right )^2 ).
\ee
Note that in this case the integral is convergent at the lower
end $s \rightarrow 0$.

\subsection*{E4-branes}

In this case the cylinder amplitude is
\be \la{e3}
Z = \int^{\infty}_{0} \frac{ds}{s} e^{i \frac{\sigma^2_6}{\pi} s},
\ee
which is clearly non convergent at both ends of the integration.
Analytic continuation can remove the $s \rightarrow \infty$
divergence, but not the one for the small $s$.
This divergence must be regulated by cutting off the integral at $s = \L$.
The regulated amplitude is thus
\be
Z = \Gamma(0, - i \L \sigma^2_6/\pi) \equiv E_1
(-i \L \sigma^2_6/\pi)
\ee
where $\Gamma(k,x)$ is the incomplete Gamma function, with
$\Gamma(0,x)$ equivalent
to the exponential integral $E_1(x)$. Expanding this for small
$x$:
\be
Z = \ln (- i \L \sigma^2_6/\pi) - \gamma +  \cdots
\ee
where $\gamma$ is the Euler constant and the ellipses denote
terms which vanish as $\sigma^2_6 \L \rightarrow 0$.

\subsection*{E6-branes}

The integral to be evaluated is
\be
Z = \int_{0}^{\infty} \frac{ds}{s}
e^{i \frac{\s^2_4}{\pi} s} (2 \sinh ( \mu X^+ s))^{-2},
\ee
which is clearly convergent at the upper end of the integration
but divergent at the lower end. In this case computing the exact
integral is difficult and thus we compute only the divergent
parts which can obtained from expanding the integrand for small $s$:
\be \la{e5}
Z = \int_{\L}^{\infty} \frac{ds}{s} e^{i \frac{\s^2_4 s}{\pi}}
(\frac{1}{4 (\mu X^+ s)^2} - {1 \over 12} + \cdots ),
\ee
where the ellipses denote finite terms. The divergent terms are thus
\bea
Z &=&
\frac{1}{(2 \mu X^+)^2} \Gamma(-2, - i \L \sigma^2_4/\pi)
- \frac{1}{6} \Gamma(0, - i \L \sigma^2_4/\pi) + \cdots \\
&=& \frac{1}{(2 \mu X^+)^2}
\left ( \frac{1}{2 \L^2} + \frac{i \s^2_4}{\pi \L} +
{\sigma^4_4 \over 2 \pi^2}
\ln( -i \L \sigma^2_4/\pi) \right )
- \frac{1}{12} \ln ( - i \L \s^2_4/\pi) + \cdots \nn
\eea

\subsubsection{Long cylinder divergences}

As is well-known, the amplitude in the long cylinder limit
should be equal to exchange diagram of massless closed
string modes. It was recently verified in \cite{DR}
that the amplitudes in the plane wave do exhibit this
behavior. Let us briefly review the field theory computation.
To compute the exchange diagram one needs the
quadratic part of the supergravity action and the
D-brane couplings. These have the form
\be
S = {1 \over 4 \k^2} \int d^{10}x \left(\psi^{\dagger}
(\Box - 2 i \mu c \pa_-) \psi
+ \delta^{10-p} (x-x_0) \l (\psi + \eta \bar{\psi}) \right)
\ee
for a Euclidean $p$-brane, where $\psi$ denotes any supergravity field
(we suppress all indices).
$(c, \l)$ are parameters that depend on the specific
gauge fixed fluctuation under consideration whilst $\eta = \pm 1$
for brane/anti-brane respectively. $c$ derives from the lightcone mass
of the fluctuation in the supergravity action and $\l$ (which is proportional
to the brane tension $T_p$) from the source in the DBI action.

The explicit constants for all cases of interest can be found
in \cite{DR}. Note in particular that there has to be a relative
factor of $i$ between NS-NS and RR field sources because the
brane is spacelike and the corresponding DBI action is analytically
continued with respect to the standard Lorentzian brane action.

Each mode of a given $(c,\l)$ contributes to the exchange between
two separated branes. and it was shown in \cite{DR} that the total
exchange was
\be
Z_{p} = - 4 \pi (4 \pi^2)^{4-p} \sin^4 (\m X^+) 
G^{10-p}(X^+,X^-,x^{r'}_1,x^{r'}_2),
\ee
where we have used the fact that $T_{p}^2 \kappa^2 = \pi (4 \pi^2)^{4-p}$
in our conventions. Here the branes have
lightcone separations $(X^+,X^-)$ and transverse positions
$(x^{r'}_1,x^{r'}_2)$ respectively.  $G^{10-p}$ 
is the propagator for a massless scalar (i.e. $\Box \psi = 0$)
over the $(10-p)$ dimensions transverse to the brane which is
given by integrating over the worldvolume directions
the $10d$ propagator \cite{Mathur}:
\be \la{10dG}
G^{10} (X^+,X^-,x^I_1,x^I_2) =
\frac{(\m X^+)^4 }{4 \pi^5 \sin^4 (\m X^+)} 
\int^{\infty}_{0} \frac{dk}{k^5} e^{\frac{i \sigma_{10}^2}{k}},
\ee
where $\sigma^2_{10}$ is the 10d geodesic separation given
in (\ref{sig}).
Integrating (\ref{10dG}) (again in the convergent regime with imaginary
$x^+$ and then analytically continuing) one gets
\be \la{10dG2}
G^{10} (X^+,X^-,x^I_1,x^I_2) = \frac{3 (\m X^+)^4 }
{2 \pi^5 \sin^4 (\m X^+)} {1 \over \sigma^8_{10}}.
\ee
Note that the limit
$\mu \rightarrow 0$ is clearly smooth, and reproduces the usual
propagator in Minkowski space.

As is very familiar,
integrating either (\ref{10dG}) (or equivalently (\ref{10dG2}))
over the first worldvolume direction in
flat space gives an overall volume factor following from translational
invariance:
\be
G^{9}_{\rm{flat}} (X^+,X^-,x_1^{r'}, x_2^{r'})
\sim \int dx^1_1 dx^1_2 \frac{1}{ (\sigma_{9}^2
+ (x_1^1 - x_2^1)^2)^4} \sim V_1 \frac{1}
{(\sigma_9^2)^{7/2}},
\ee
where $V_1$ is the regulated length and $\sigma_9^2$ is the 9d
geodesic separation. Thus for the field theory exchange between spacelike
p-branes in flat space one finds the usual
\be \la{ex1}
Z_{{\rm{flat}}} \sim \frac{V_{p}} { (\sigma_{10-p}^2)^{(8-p)/2}},
\ee
where the overall prefactor is of course zero for brane/brane
field exchange because of supersymmetric cancellation.

In the plane wave, translations in the $x^I$ directions act as
\be \label{xitr}
\d x^- = \m \sin \m x^+ \e^I x^I, \qquad \d x^I = \cos \m x^+ \e^I.
\ee
This implies that the system of branes separated along the lightcone
directions
is not generically invariant under translations in the $x^I$ directions
and it leads to very different behavior for the integrated propagators compared
to flat space: there is no overall volume factor and the power law behavior is
modified. Thus
\be \la{zp}
Z_p = - 2^{8-2p} (\pi \mu X^+)^{4-p}
\int_{0}^{\infty} \frac{dk}{ k^{5-p}} e^{i \frac{\s^2_{10-p}}{k}},
\ee
which clearly reproduces the small $s$ or equivalently large $t$
behavior of the integrands in the string amplitudes (\ref{ca}) and
(\ref{oa}) for $p=2$ and $p=4$. For the E2-brane one gets
the finite answer given in \cite{DR}
\be
Z_{p=2} = - 2^4 \frac{
(\pi \mu X^+)^2}{(\sigma_8)^4},
\ee
in agreement with the
large $\sigma^2$ behavior of the string amplitude.
For $p=4$ the field theory amplitude is clearly exactly
the string amplitude (\ref{e3}) and reproduces its logarithmic
divergence.

For $p=6$ the expression (\ref{zp})
is no longer valid: (\ref{zp}) was obtained by exchanging
the order for the integrations over $k$ and $x^r$ respectively which
is only permitted when the integrals are convergent. However,
one can straightforwardly integrate (\ref{10dG2}) over the
worldvolume coordinates
to reproduce the divergent parts of the string amplitude given
in (\ref{e5}).

\bigskip

We have been discussing the field theory exchange between
two parallel separated branes. In flat space the long range
supergravity fields sourced by a single brane are translationally
invariant along the Neumann directions and are proportional to
the relevant propagator $1/\sigma_{10-p}^{8-p}$, i.e. (\ref{ex1})
without the overall volume factor.

In the plane wave the absence of translational invariance means
that the long range supergravity field sourced by the brane depends
on the position in the Neumann directions. The explicit behavior is given
by integrating the propagator over the worldvolume directions. For
the behavior of a massless supergravity mode $\psi$ (i.e. $c = 0$)
at a given point far from the brane one gets
\be
\psi \sim (\mu X^+)^{4 -\half p} \frac{\tan^{\half p} (\mu X^+)}
{\sin^4 ( \mu  X^+)} \hat{\s}^{p-8},
\ee
where
\bea
\hat{\s}^2 &=& 2 X^+ X^- - \mu X^+ \tan(\mu X^+) \sum_{r} (X^r)^2 \\
&& + \frac{\mu X^+}{\sin(\mu X^+)}
\sum_{r'} \left ( ((x_b^{r'})^2 + + (X^{r'})^2 )\cos(\mu X^+)
- 2 x^{r'}_b X^{r'} \right ), \nn
\eea
and $x^{r'}_b$ is the brane position, $X^I$ is the transverse position
of the observation point and $(X^+,X^-)$ is the lightcone separation between
the brane and the observation point. Thus the power law dependence is
the same as in flat space but the field sourced is not
translationally invariant
along the directions parallel to the brane. Integrating with respect to
$X^r$ gives the brane/brane exchange behavior, again
demonstrating that the logarithmic and power law divergences discussed above
result from the infinite volumes of the branes.

\subsection{Distinguished $x^+$ separations and plane wave geodesics}

In the previous section we have discussed features of the
amplitudes for generic separations, emphasizing the lack of
translational invariance. For special brane separations, however,
translational invariance is restored. This happens when
\be \label{sep}
\m X^+ = l \pi, \quad \ l \in Z
\ee
In this case, (\ref{xitr}) yields $\d x^-=0,\ \d x^I = (-1)^l \e^I$
and one expects the amplitudes to become similar to the flat space
amplitudes.

Physically one can understand these distinguished values as arising from
the behavior of geodesics in the plane wave:
a generic geodesic will reconverge to the same transverse position $x^I$
after evolution by
$\mu X^+ = 2 \pi$. Labelling the geodesic by $X^+$, its trajectory is
\cite{BMZ,Mathur, Dorn}
\bea
x^- (X^+) &=& x_1^- + \qu (x_1^2 - p_1^2) \sin (2 \mu X^+) - \half x_1
\cdot p_1 \cos (2 \mu X^+)
+ C X^+ + \half x_1 \cdot p_1; \nn \\
x^{I} (X^+) &=& x^I_1 \cos ( \mu X^+) + p^{I}_1 \sin ( \mu X^+),
\eea
where $(x^{-}_1, x^I_1, p_1^I)$ are initial conditions for the geodesic.
The constant $C$ is also given in terms of initial conditions
via $p_{1}^- + \half \mu (p_1^2 - x_1^2)$.

Thus a generic geodesic will reconverge to its original transverse
position $x^{I}_1$ after $\mu X^+ = \pi l$ with $l$ even and it
will pass through $ - x^{I}_1$ for $l$ odd. After evolution
through $\mu X^+ = \pi l$, the geodesic will be shifted in $x^-$
by an amount $\pi l C/\mu$. Such focusing is unavoidable given finite valued
initial conditions for the geodesic; one can only avoid
the focusing with infinite initial velocities for the geodesic (i.e.
$p_1$ is infinite).

The focusing of geodesics can be understood
in terms of focusing of geodesics on $AdS_5 \times S^5$. As was reviewed
in section 2, there is a focusing of geodesics between the north and 
south poles
of the $S^5$ and after $t=\pi$ on global $AdS_5$. Now, recall that in 
taking the Penrose limit one defines coordinates
\be
\mu x^+ = \half (\t + \q), \qquad x^- = {\m R^2 \over 4} (\q - \t)
\ee
and then takes the limit $R^2 \to \infty$. Clearly for $x^-$ to stay finite in the limit
we need
\be
\q = \tau + {\cal O}({1 \over R^2})
\ee
so the geodesics that join focal points of $AdS_5$ {\it and} of the circle of $S^5$ along
which we boost survive in the limit. These are the geodesics discussed above.

Note that (\ref{sig}) implies
that the geodesic distance becomes infinite for $\mu X^+ = l \pi$
if $(x_1^I + (-1)^{l+1} x_2^I) \neq 0$ even if the latter is finite.
To regulate the geodesic distance we consider the $x^+$ separation
to be given by
\be
\m X^+ = l \pi+\e, \ l \in Z
\ee
where $\e$ is infinitesimal. The geodesic distance becomes
\be
\s_{10}^2 = \sum_{I=1}^8 \left (
{\pi l \over \e} (x_1^I + (-1)^{l+1} x_2^I)^2 +
( \frac{2 \pi l X^-}{\mu} + (x_1^I + (-1)^{l+1} x_2^I)^2)  \right )
+ {\cal O} (\e),
\ee
which clearly shares the translational invariance of the flat space geodesic
distance. Now consider the behavior of the massless field 
propagator: from (\ref{10dG2}) one sees that it is finite as 
$\e  \rightarrow 0$, and is exactly the same as in flat space. 
Integrating over the worldvolume coordinates
now yields the following expression for the total field theory exchange
\be \la{rfte}
Z_p \sim -
l^{4 - \half p} V_p
\ep^{\half p} \int^{\infty}_{0} \frac{dk}
{k^{5 - \half p}} e^{i \frac{\s^2_{10-p}}{k}}
\sim -
{V_p \ep^4 \over ( \sum_{r'} (x_1^{r'} 
+ (-1)^{l+1} x_2^{r'})^2 )^{4- \half p}},
\ee
where $V_p$ is the regulated brane volume and overall numerical
factors are suppressed. This expression
explicitly demonstrates the reinstated translational invariance.
Moreover the total amplitude for field theory exchange 
vanishes as $\ep \rightarrow 0$.

The above implicitly assumes that the brane separations
in the $x^{r'}$ Dirichlet
directions are such that the geodesic distance between the branes diverges
in the limit $\m X^+ = l \pi$. When the brane positions take
the special values $x_1^{r'} = (-)^l x_2^{r'}$ the geodesic separation
remains finite in this limit. As we have discussed, for these
separations there are an infinite number of geodesics
connecting the two branes along each such Dirichlet direction rather
than a unique geodesic as is usually the case.

Suppose that $x_1^{r'} = (-)^l x_2^{r'}$ for $j$ of the Dirichlet directions,
i.e. the branes are either coincident in these directions or their positions
are reflections of one another; the amplitude in this limit can conveniently 
be obtained by taking the limit of the exponential
in the integrand in (\ref{rfte}) as $x_{1}^{r'} \rightarrow (-)^l x_{2}^{r'}$
using the identity (here and in subsequent expressions in this
section integrals are implicitly computed in the convergent
regime and then analytically continued; overall phase factors are
suppressed)
\be
\d(x) = \lim_{\a \rightarrow \infty} (\frac{\a}{\sqrt{\pi}}
e^{- \a^2 x^2}),
\ee
along with $2 \pi \d(x=0) = \td{V}$, where $\td{V}$ is the volume
of momentum space\footnote{This can be seen using $\d(x)=\int dp/(2 \pi) e^{i p x}$, so $2 \pi \d(0) = \int d p$.}.
The amplitude then becomes
\be \la{d1}
Z_{p} \sim - { V_p \td{V}_j \ep^4 \over (\sum_{r' = p + j +1}^{8-p}
(x_1^{r'} + (-)^{l+1} x_2^{r'})^2)^{4 - \half (p+q)}},
\ee
where $\td{V}_j$ is the regulated momentum space volume for the $j$
Dirichlet directions.
Note that this expression is valid only for $j < (8 - p)$; in the limit
$j = (8-p)$ one obtains
\be \la{d2}
Z_p \sim V_{p} \td{V}_{8-p} \ep^4 \int^{1/\Lambda}_{0} \frac{dk}{k}
e^{ \frac{2 \pi i l X^-}{\mu k }} \sim V_p \td{V}_{8-p} \ep^4
\ln ( l X^- \Lambda/\mu),
\ee
where $\Lambda$ is a regulator.

The key features of both expressions (\ref{d1})
and (\ref{d2}) are that the amplitude scales with the Dirichlet volume
and vanishes in the limit $\ep \rightarrow 0$.
We now consider these distinguished separations
in the open and closed string channels; this will clarify the
physical origin of both these features.

\subsubsection{Open string channel}

Let us consider the limit $\mu r^+ \rightarrow 1$
in the open string mode expansions given previously. For notational
ease we discuss the specific case $l =1$ but
general $l$ follows straightforwardly from this case.
In this limit the frequencies of the first stringy modes approach zero
and one finds that
\bea
x^{r}(\s,\t) &=& \frac{1}{\sqrt{2}}
(X_0^r + P_{0}^r \t) \cos (\s) + \cdots \label{neu1} \\
x^{r'}(\s,\t) &=& x_{1}^{r'} \cos (\s) +
\frac{1}{\sqrt{2}} (X^{r'}_0 + P^{r'}_0 \t) \sin(\s) + \cdots \label{dir1} \\
\sqrt{r^+} \q^1 &=& (M \Pi \cos(\s) - i \sin (\s)) (\q_1 - \q_{-1})
+ \cdots \la{fer1} \\
\sqrt{r^+} \q^2 &=& (M^t \sin(\s) - i \Pi \cos(\s)) (\q_1 - \q_{-1})
+ \cdots \la{fer2}
\eea
where the ellipses denote unaffected terms in the mode expansions
(i.e. $n^2 \neq 1$). The associated commutation relations are
\be
[ X_{0}^{r}, P_{0}^{s} ] = i \d^{rs}, \hsp
[ X_{0}^{r'}, P_{0}^{s'} ] = i \d^{r's'}, \hsp
\{ \q_{1}, \q_{-1} \} = \qu.
\ee
The Dirichlet solution is rather special, in that one
loses an integration constant in this limit because the second
``zero mode'' solution coincides with the limit of the stringy mode
solution. The string is forced to have its endpoints at
$\pm x^{r'}_1$; this fits with the geodesic behavior, in that
only infinite proper length geodesics will give $x_2^{r'}
\neq - x_1^{r'}$. We can thus only consider this latter case when
the geodesic distance is regulated via $\td{m} = (1 + \ep/\pi)$ as in the
previous discussion.

Computing the contributions to the
conserved charge $\hat{H}_o$ from these modes one finds
\be
\hat{H}_o =  i (\sum_{r=1}^p a_0^r \bar{a}_0^r - 2 i \q_0 M \Pi \q_0
 + \half p)
+ \half \sum_{I} (P_0^I)^2 + \sum_{n > 1} \w_n (a^I_{-n} a^I_n + 4
\q_{n} \q_{-n}).
\ee
Thus fermion modes $\q_{\pm1}$ drop out of the charge
in this limit since their frequencies are zero. States
can now be labelled simultaneously by their
${\hat{H}}$ eigenvalue and by their continuous ``momentum''
$P_0^I$. Computing the annulus using this Hamiltonian we obtain
\be \la{x+a}
Z =  \frac{V_{p} V_{8-p}}{(2 \pi)^8}
\int \frac{ds}{s^5}  e^{2 i X^- s/\mu}
(2 \sinh(\pi s))^{4-p} (1 - 1)^8,
\ee
where $V_{p}$ and $V_{8-p}$ are the regulated volumes of
the Neumann and Dirichlet directions respectively,
the volume factors originating from the standard identity
${\rm{Tr}} (e^{- \pi P^2 s}) = V/(4 \pi^2 s)^{\half}$. Note
that the volume appearing here is the position space volume.
The unbalanced massive zero mode harmonic oscillators give the same
contribution as for generic $X^+$ but there is
now an overall factor of $(1 - 1)^8$ from the massless fermionic modes,
which annihilates the amplitude.

The existence of continuous modes in the Neumann
directions reflects the reinstated translational invariance
along the worldvolume already noted and leads to the $V_p$ factor.
There are also continuous modes in
the Dirichlet directions which leads to the $V_{8-p}$ volume factor;
from the mode expansions, one can see that these
follow directly from the infinite family of geodesics connecting
$x^{r'}$ and $-x^{r'}$ when $\mu X^+ = \pi$.

Note that the scaling of the amplitude with the volume is
also a feature of the brane/antibrane amplitude, which can be shown
to be
\be \la{babr}
Z = \frac{V_{p} V_{8-p}}{(2 \pi)^8}
\int \frac{ds}{s^5}  e^{2 i X^- s/\mu} (2 \sinh(\pi s))^{4-p}
\left ( { f_4^1(\td{q}) \over \td{h}^1_1(\td{q})} \right )^8,
\ee
where the function $\td{h}^1_1(\td{q})$ is related to the modular
function $f_1^{\td{m}}(\td{q})$ as follows
\be
\lim_{\td{m} \rightarrow 1 + \ep/\pi} f_1^{\td{m}}(\td{q}) =
\sqrt{8 \pi \ep} s \td{h}^1_1(\td{q}) + {\cal O}(\ep)
\ee
and is given by
\be
\td{h}^1_1(\td{q}) = \td{q}^{- \D_{1}} (1 - \td{q}^{i})^{\half}
\prod_{n=2}^{\infty}(1 - \td{q}^{\sqrt{n^2 -1}}).
\ee
The modular function $f^1_4(\td{q})$ is given by the modular function of 
\cite{BGG}
\be
f_4^{\td{m}}(\td{q}) = \td{q}^{-\D_{\td{m}}'} \prod_{n=1}^{\infty} (1 - \td{q}^{\sqrt{(n - \half)^2
- \td{m^2}}})
\ee
in the particular limit $\td{m} \rightarrow 1$.

\bigskip

Now consider the behavior of the brane/brane amplitude (\ref{onel})
computed for generic separations as one takes the limit 
$\td{m} = 1 + \ep/\pi$.
The relevant behavior is clearly that of the first stringy modes and thus
\be
Z \sim \la{lim}
\ep^{\frac{1}{2} (p+j)}
V_p V_j \int^{\infty}_{0} \frac{ds}{s} e^{i \frac{\sigma^2_{10-p-j}}
{\pi} s} (2 \sinh (\pi s))^{4-p} s^{\frac{1}{2} (p+j)}
\left ( \frac{s \td{h}^1_1(\td{q}) } {s \td{h}^1_1(\td{q})} \right )^8,
\ee
where $x_1^{r'} = - x_2^{r'}$ for $j$ of the Dirichlet directions.
(We left the factors of $s$ with $\td{h}^1_1$ for later convenience).
The volume factors in the Neumann and Dirichlet directions arise
from the limiting behavior of the first stringy modes, namely
\bea
N &:& \lim_{\w_1 \rightarrow 0} {\rm{Tr}}e^{-2 \pi i s \w_1 a_1 a_{-1}}
\rightarrow {\rm {Tr}} e^{- i \pi P_0^2}
\sim {V \over \sqrt{s}}; \\
D &:& \lim_{\w_1 \rightarrow 0} \left( \lim_{x_1 \to - x_2}
\left ( e^{i (x_1 + x_2)^2 s/\ep} {\rm{Tr}}
e^{- 2 \pi i s \w_1 a_1 a_{-1}} \right )\right)
\rightarrow {\rm {Tr}} e^{ - i \pi P_0^2}  \sim {V \over \sqrt{s}}. \nn
\eea
The corresponding amplitude for brane/anti-brane is given by replacing
the factors of $s \td{h}^1_1(\td{q})$ in the numerator by
$f_4^1(\td{q})/\sqrt{\ep}$; the result for the limiting behavior of the
amplitude then clearly agrees with (\ref{babr}). Comparison
of the brane/brane amplitudes (\ref{x+a}) and (\ref{lim})
is slightly less clear, since they both tend to zero in
this limit; they do however agree if one identifies
the massless fermion contributions via $s^8 \ep^4 \rightarrow
(1 - 1)^8$.

\subsubsection{Closed string channel}

Distinguished values of $X^+$ are also visible in the closed
string description. Looking at the annulus in the closed
channel (\ref{ca}), however, it is apparent that the
special behavior derives from the zero (supergravity) modes,
in contrast to the open channel where for $\mu X^+ \rightarrow l \pi$
the $l$th stringy modes become massless. The amplitude (\ref{ca}) in the
limit $\mu X^+ = l \pi + \ep$ becomes
\be \la{amp}
{\cal A} \sim V_{p} \td{V}_j \ep^{\half (p+j)} l^{\half (p-j)}
\int_{0}^{\infty} \frac{dt}{t^{1 + \half (p-j)} }
e^{ \frac{i \s^2_{10-p-j}} {\pi t}} (\sinh (\frac{\pi l}{t}) )^{4- p}
\left ( \frac{ h^l_1(q)}{ h^l_1(q)} \right )^8,
\ee
where the function $h^l_1(q)$ is defined by the limit
\be
\lim_{m \rightarrow (l+\ep/\pi)/t} f_1^m(q) =
\sqrt{2 i \ep} h^l_1(q) + {\cal O}(\ep)
\ee
and is given by
\be
h^l_1(q) = q^{-\D_{l/t}} \prod_{n=1}^{\infty} (1 - q^{\sqrt{n^2
+ l^2/t^2}}).
\ee
In the amplitude (\ref{amp}) the brane separations are again such
that $x_1^{r'} = (-)^l x_2^{r'}$ for $j$ of the Dirichlet directions
and $V$ and $\td{V}$ denote position space and momentum space volumes
respectively. In this case the volume factors arise from the limits
\bea
N &:& \lim_{p \rightarrow 0} \left(\lim_{mt \rightarrow l}
 \langle -p | e^{-2 \pi i t H} | p \rangle \right)
\sim \lim_{p \rightarrow 0} \left(\lim_{\ep \rightarrow 0}
\frac{1}{\sqrt{\ep}} e^{2 i l p t \over \e } \right)
\sim l^{\half} t^{- \half} {V}; \\
D &:&  \lim_{x_1 \rightarrow (-)^l x_2} \left(\lim_{mt \rightarrow l}
 \langle x_1 | e^{-2 \pi i t H} | x_2 \rangle \right)
\sim \lim_{x_1 \rightarrow (-)^l x_2} \left(\lim_{\ep \rightarrow 0}
\frac{1}{\sqrt{\ep}} e^{i l (x_1 -(-)^l x_2) \over \e t} \right)
\sim l^{- \half} t^{\half} \td{V}. \nn
\eea
The amplitude reproduces in the large $t$
long cylinder limit the previous expression for the supergravity
field theory exchange (\ref{d1}).

The open/closed duality between the two expressions (\ref{lim}) and
(\ref{amp}) does not follow trivially from the previous proof for generic
brane separations. The $h$ functions inherit modular transformation
properties from the $f$ functions, namely $h^1_1(q) = s \td{h}^1_1(\td{q})$,
which ensures that the Neumann and fermion contributions to
the integrands in (\ref{lim}) and (\ref{amp}) agree upon modular
transformation. For the Dirichlet modes, however, there is a subtlety in
that the volume appearing in (\ref{lim}) is that of $X_0^{r'}$
whilst that in (\ref{amp}) is that of $p_0^{r'}$. There is
a simple way to relate these quantities. Consider a cylindrical
worldsheet. In the open channel the Dirichlet zero modes
on the cylinder are
\be
x^{r'}(\s) = x_1^{r'} \cos(\s) + \frac{1}{\sqrt{2}} X_0^{r'} \sin(\s).
\ee
Note that the $P_0^{r'} \t$ modes present for the tree level open strings
are absent on the cylinder since they are not consistent with periodicity in
time; this is the origin of the $\d(P_0^{r'} = 0) \sim V_{X_0^{r'}}$ factor
in the annulus. In the closed channel the Dirichlet zero modes are
\be
x^{r'}(\t) = x_0^{r'} \cos(m \t) + m^{-1} p_0^{r'} \sin(m \t).
\ee
Under open/closed duality the two sides of the cylinder are exchanged,
$\s \leftrightarrow m \t$, as discussed earlier.
The matching of the cylinders in the two channels thus requires the
identification $X_0^{r'} \sim m^{-1} p_0^{r'} \sim t p_0^{r'}$
and thus $V_{j} \sim t^{j} \td{V}_j$. This identification ensures
that the Dirichlet contributions to the amplitudes (\ref{lim})
and (\ref{amp}) agree upon modular transformation.

\bigskip

The distinguished brane separations are also immediately apparent
when one considers the Hamiltonian
evolution of the boundary state. A boundary state at general $x_0^+$
can be obtained from that at $x^+_0 = 0$ by acting with the evolution
operator $e^{ -i H x_{0}^+/p^+} \equiv e^{-i H \t_0}$.
It can equivalently be obtained
by writing down gluing conditions for a worldsheet boundary at
$- \t_0$. The latter are
\bea \la{bc1}
&& \hspace{-.7cm}
(p_{0}^{r} + m x_{0}^r \tan (m \t_0)) \bol_{\t_0} = 0; \hsp
 (\a^{1r}_{n} + e^{-2 i \w_n \t_0} \a^{2r}_{-n}) \bol_{\t_0} = 0; \\
&& \hspace{-.7cm}
(x_{0}^{r'} - m^{-1} \tan(m \t_0) p_0^{r'} - \frac{x^{r'}}
{\cos(m \t_0)} ) \bol_{\t_0} = 0; \hsp
(\a^{1r'}_{n} - e^{-2 i \w_n \t_0} \a^{2r'}_{-n}) \bol_{\t_0} = 0, \nn \\
&& \hspace{-.7cm}
(\cos (2 m \t_0) \q_{0}^1 + (i \eta M - \Pi \sin (2 m \t_0) )
\q_{0}^2) \bol_{\t_0} =
(\q_{-n}^{1} + i \eta M e^{-2 i \w_n \t_0} \q^2_n) \bol_{\t_0} = 0. \nn
\eea
From these expressions one sees that the action of the time evolution
on the stringy modes is rather unimportant, a phase rotation, so
the boundary state is
\be
\bol_{\t_0} = {\cal{N}} \exp \left ( \sum_{n=1}^{\infty}
e^{-2 i \w_n \t_0}
(\w_n^{-1} M_{IJ} \a^{I1}_{-n} \a^{J2}_{-n} - i \eta M \q^1_{-n}
\q^2_{-n}) \right ) \bo1_{\t_0},
\ee
where $\bo1_{\t_0}$ is the zero mode part.
The explicit solution for the bosonic zero mode part of
the boundary state is
\be
\exp ( - \half \sum_r e^{-2 i \mu x_0^+} a_0^r a_0^r
+ \half \sum_{r'} (e^{-i \mu x^+_0} a_{0}^{r'}
- i \sqrt{2m} \cos (\mu x_0^+) x^{r'} )^2) \vac
\ee
whilst following (\ref{vac1}), (\ref{vac2}) the fermion
zero mode part of the boundary state for E2 and E6 branes
(for which $M^2 = -1$) is given by
\be
\exp ( - \half \eta ( e^{2 i \Pi \mu x^+_0} M)_{a b}
\q_{L}^a \q_L^b + \half \eta (e^{- 2 i \Pi \mu x^+_0} M)_{ab}
\bar{\q}_R^a \bar{\q}_R^b) \vac,
\ee
whilst that for E4 branes is slightly different
but analogous (since in this case $M^2 = 1$).

The physical interpretation is the following.
The lightcone Hamiltonian describes the past and future evolution of
a boundary state defined at some given $x_0^+$, which can be fixed
to zero via translational invariance. Initially the Neumann and
Dirichlet boundary conditions, $ \pa_{\t} X^r = 0$ and $X^{r'} = x^{r'}$
respectively, imply that the state is of zero momentum in the $r$ directions
and at fixed position in the $r'$ directions.

As the state evolves in $x^+$,
however, the source effectively rotates in the $x^I$ directions, the time evolution
reflecting the behavior of the geodesics. Consider a Neumann direction: the boundary state is
a zero momentum eigenstate, as is usual for a Neumann direction,
for $\mu x^+_0 = l \pi$ but a zero position eigenstate for
$\mu x^+ = l \pi/2$ for $l$ odd. In between it is a mixed eigenstate,
neither pure position nor pure momentum.
Similarly the Dirichlet directions are pure position eigenstates
at $\mu x^+_0 = l \pi$ but pure momentum eigenstates at
$\mu x^+_0 = l \pi/2$ for
$l$ odd and mixed eigenstates in between these values.
There is an analogous periodicity in the fermion zero modes:
the boundary state is annihilated by the same combination of
zero modes after evolution through $\mu x_{0}^+ = l \pi$
and by precisely the opposite combination of zero modes
after evolution through $\mu x_{0}^+ = l \pi/2$ with $l$ odd.

Thus there is effectively a worldvolume transmutation. Take
for instance the case of $(2,0)$ branes. After evolution through
$\mu x^+ = l \pi/2$ with $l$ odd the source becomes localized in the $(1,2)$
directions but uniformly distributed over the other 6 directions
transverse to the lightcone: the brane is effectively an E6 brane. Note
however that the evolved state for the E2 brane coincides with the initial
state for the E6 brane only in zero modes; the stringy mode parts are
different. Similarly the $(3,1)$ E4 brane after evolution by $\mu x^+ = l
\pi/2$ becomes a $(1,3)$ E4 brane.

One should contrast this behavior with that of the analogous branes
in flat space: taking $m \rightarrow 0$ in (\ref{bc1}) one sees
that the Neumann and fermion zero mode conditions are independent
of $\t_0$, since they commute with the Hamiltonian, but the Dirichlet
zero mode condition gives
\be
(x_0^{r'} - \t_0 p_0^{r'}  - x^{r'}) \bol_{\t_0} = 0,
\ee
which describes a position eigenstate at $\t_0 = 0$ but a zero momentum
eigenstate as $\t_{0} \rightarrow \pm \infty$. Thus an initial
source localized at some Dirichlet position totally disperses
over the Dirichlet directions in the far future and past. The difference
in the plane wave is that the effective harmonic oscillator potential
prevents the source from dissipating, and causes it to recollapse
at periodic intervals in $x^+$.
It would be interesting to analyze whether these branes present
interesting cosmological models for cyclic universes in the context of braneworld scenarios.

\subsection{Comments on Lorentzian branes}

One might wonder why the spacelike branes in the plane wave
rather than the usual Lorentzian branes have been used to illustrate
generic properties of branes in curved backgrounds. The reason
is that the properties under discussion here
are only visible in the plane wave
for objects separated in the $x^+$ direction. The amplitudes
for Lorentzian branes at leading order are the same as those for
branes in flat space \cite{Hammou,DR}.
Let us briefly review this argument. (See also \cite{Durin}
for a related discussion for the annulus of
branes in flat space carrying traveling waves.)
The cylinder amplitude between parallel separated Lorentzian branes is
\be
Z = V_{+-} \int_{0}^{\infty} \frac{ds}{s} \int dp^+ dp^-
{\rm{Tr}} ( (-)^F e^{- i (p^+ p^- + H) s} ),
\ee
where $V_{+-}$ is the regulated volume of the lightcone and $H$ is the
Hamiltonian for a Lorentzian Dp brane in standard lightcone
gauge $x^+ = p^+ \t$.
Carrying out the $p^-$ integration gives $\d(p^+ s)$ which enforces
the limit $H_{p^+ \rightarrow 0}$ i.e. all $m$ dependence drops out
and the Hamiltonian is the same as in flat
space. Thence the overall amplitude is exactly as in flat space, zero
because of the (now massless) fermion zero modes.

In particular, the cylinder amplitude vanishes even for branes
displaced from the origin in the plane wave. Such branes admit
dynamical supersymmetries in their spectra which are not
expected to be preserved by interactions \cite{FST}. The
supersymmetries of the spectra along with the projection onto
$p^+ \rightarrow 0$ lead to the vanishing of the cylinder amplitude.
It is possible, however, that the self amplitude for these branes
at the next order ($g_s$) is non-trivial and that it develops an imaginary
part corresponding to the decay of these branes, presumably back
to the origin in the plane wave. Even if
this is the case, these branes are certainly stable in perturbation
theory because their decay time is at least of order $1/g_s$.

One should take with some caution the arguments given above for
the vanishing of the cylinder, since the integral has projected onto
states with $p^+ = 0$ which are of course precisely those which
are inaccessible in lightcone gauge. However, independent confirmation
for these arguments comes from considering the field theory limit
of the exchange. Following the arguments of the previous sections,
one needs to integrate the 10d propagator over the worldvolume
directions, which now include the lightcone. Integrating the propagator
over $x^-$ clearly projects onto $x^+ = 0$, in which limit one recovers
exactly the flat space behavior: the translational invariance over
the $x^I$ directions is reinstated and the overall amplitude vanishes.

\section*{Acknowledgments}

MT would like to thank the Benasque Center for Science and the
Aspen Center for Physics for hospitality during the completion
of this work. KS is supported by NWO.

\appendix

\section{Closed string mode expansions}

Given the lightcone gauge fixed action (\ref{sp+})
the closed string mode expansions are given by
\bea
x^{I}(\s,\t) &=& \cos (m \t) x_{0}^I + m^{-1}
\sin (m\t) p_{0}^{I} + i \sum_{n \neq 0} \frac{1}{\sqrt{2}}
\w_{n}^{-1} (\a_n^{1 I}
\td{\f}_n + \a_n^{2I} \f_n); \\
\sqrt{2 p^+} \q^1 (\s,\t) &=&
\q^1_0 \cos (m\t) + \Pi {\q}^2_{0} \sin (m\t)
+ \sum_{n \neq 0} c_n \left ( i d_{n}\Pi \q^2_n \phi_n
+ {\q}^1_{n} \td{\f}_{n} \right ) ; \label{fercl} \\
\sqrt{2 p^+} \q^2 (\s,\t) &=& {\q}^2_0 \cos (m\t) - \Pi {\q}^1_{0} \sin (m\t)
+ \sum_{n \neq 0} c_n \left ( - i d_n \Pi {\q}^1_n \td{\f}_n
+ {\q}^2_{n} \f_{n} \right ),
\eea
where the expansion functions are
\be \la{fn}
\f_n(\t,\s) = e^{-i(w_n \t + n \s)}, \qquad
\tilde{\f}_n(\t,\s) = e^{-i(w_n \t - n \s)}.
\ee
After canonical quantization we get the following (anti)commutators
\bea
[p_{0}^{I},x_{0}^{J}] = - i \d^{IJ}, \hsp
[\a_{m}^{\ca I},\a_{n}^{\caj J}] = \w_{m} \d_{n+m,0}
\d^{\ca \caj} \d^{IJ}, \\
\{ \q_{0}^{\ca a }, \q_{0}^{\caj b } \} = \d^{\ca \caj} \d^{ab},
\hsp
\{ \q_{m}^{\ca a }, \q_{n}^{\caj b } \} =
\d^{\ca \caj} \d_{m+n,0} \d^{ab}, \label{clcm}
\eea
where $\ca = 1,2$.
It is convenient to introduce creation and annihilation operators
\be
a_{0}^{I} = \frac{1}{\sqrt{2m}} (p_0^{I} + i mx_{0}^{I}), \hsp
\bar{a}_{0}^{I} = \frac{1}{\sqrt{2m}} (p_0^{I} - i m x_{0}^I), \hsp
[\bar{a}_0^{I},a_{0}^J] = \d^{IJ}.
\ee

\section{Evaluation of Dirichlet zero mode amplitude}

In this appendix we discuss the evaluation of the Dirichlet
zero mode part of the amplitude:
\be \la{dzm}
{\cA}_{D} =\langle 0 | e^{\frac{1}{2} (\bar{a}_0 + i
\sqrt{2 y_1})^2} e^{\ln(z) a_{0} \bar{a}_0}
e^{\frac{1}{2} (a_0 - i \sqrt{2 y_2})^2} | 0 \rangle,
\ee
where $y_i^{r'} = m (x_i^{r'})^2, i=1,2$.
To evaluate this we make use of the Campbell-Baker-Hausdorff
formulae
\be \la{cbh}
e^{P} e^{Q} = e^{P + \cL_{\hp P} (Q +
\coth(\cL_{\hp P}) Q) + ...} =
e^{Q + \cL_{\hp Q} ( -P  + \coth(\cL_{\hp Q}) P) + ...},
\ee
which contain all terms linear in the operator Q (P) in
the first (second) formula. The Lie derivative is defined as
\be
\cL_{\hp P} Q = \half [P, Q],
\ee
and the hyperbolic cotangent should be evaluated as a power series
expansion in
\be
\cL_{\hp P}^{n} Q = \left [\half P, \left [ \half
P, ... [\half P,Q] \right ] \right ],
\ee
with $n$ factors $\hp P$. {\footnote{Recall that the expansion of the
hyperbolic cotangent is
$x \coth(x) = \sum_{n=0}^{\infty} \frac{2^{2n}}{(2n)!} B_{2n} x^{2n}$
where $B_{2n}$ are Bernoulli numbers.}}
The formulae (\ref{cbh}) are exact provided that all commutators involving
more than one Q (P) vanish. Using these formulae one can show that
\be
e^{ \ln(z) a_{0}
\bar{a}_0} e^{\hp (a_{0} - i \sqrt{2 y_2})^2 } | 0 \rangle
= e^{\hp (z a_{0} - i \sqrt{2 y_2})^2 } e^{
\ln(z)  a_0 \bar{a}_0} | 0 \rangle =
e^{\hp (z a_{0} - i \sqrt{2 y_2})^2 } | 0 \rangle,
\ee
since $\bar{a}_{0} \vac = 0$. Thus (\ref{dzm}) reduces to
\be \la{e1}
\cA_D (z,y_1,y_2) =
\langle 0 | e^{\hp (\bar{a}_0 + i \sqrt{2 y_1})^2}
e^{\hp (z a_{0} - i \sqrt{2 y_2})^2 } | 0 \rangle.
\ee
The easiest way to compute this amplitude is to show that it satisfies a
set of differential equations which can then be integrated.
For instance, setting $y_1=y_2=0$ and differentiating w.r.t. $z$
yields
\be
\pa_z \cA_D (y_1{=}y_2{=}0) = z B
\ee
where $B=\langle 0|e^{\half \bar{a}_0^2} a_0^2 e^{\half z^2 a_0^2} |0 \rangle$.
Commuting $a_0^2$ to the left where it annihilates $ \langle 0|$ yields
$B = \cA_D (y_1{=}y_2{=}0)+z^2 B$ and we finally arrive at
\be
\pa_z \cA_D (y_1{=}y_2{=}0) = \frac{z}{1-z^2} \cA_{D}(y_1{=}y_2{=}0).
\ee
Similar manipulations yield the following differential equations
\bea \la{dif1}
\pa_{y_1} \cA_D &=& \frac{1}{1-z^2} ( z \sqrt{\frac{y_2 }{y_1}} - 1)
\cA_D; \\
\pa_{y_2} \cA_D &=& \frac{1}{1-z^2} ( z \sqrt{\frac{y_1}{y_2}} -
1) \cA_D. \nn
\eea
These equations together with $\cA_{D} (z{=}y_1{=}y_2{=}0)=1$
imply
\be
\cA_D = \frac{1}{(1-z^2)^{\hp}} e^{ - \frac{1}{1-z^2} (y_1 + y_2 -
2 z \sqrt{y_1 y_2})}.
\ee

\end{document}